\documentclass[preprint,pdf]{iucr}              

                   \def\href#1{\relax}\let\foo\caption
\ifPDF
  \RequirePackage{hyperref}
  \PassOptionsToPackage{pdftex,bookmarksopen,bookmarksnumbered}{hyperref}
\fi
\let\caption\foo

    \paperprodcode{a000000}      
    \paperref{xx9999}            
    \papertype{FA}               
    \paperlang{english}          
    \journalcode{A}              

    \journalyr{2000}
    \journaliss{1}
    \journalvol{56}
    \journalfirstpage{000}
    \journallastpage{000}
    \journalreceived{0 XXXXXXX 0000}
    \journalaccepted{0 XXXXXXX 0000}
    \journalonline{0 XXXXXXX 0000}

\usepackage[utf8]{inputenc}

\usepackage{hyperref}
\usepackage{bookmark}

\usepackage{amsmath, amssymb}
\usepackage{mathtools}
\usepackage{bm}

\usepackage{ascmac}

\usepackage{url}

\usepackage{graphicx}

\usepackage{tikz}
\usetikzlibrary{positioning,arrows.meta,shapes}

\usepackage{tabularx}
\newcolumntype{Y}{>{\centering\arraybackslash}X}
\newcolumntype{b}{Y}
\newcolumntype{s}{>{\hsize=.5\hsize}Y}

\newcommand{\norm}[1]{\left\lVert#1\right\rVert}
\DeclareMathOperator*{\argmin}{arg\,min}
\DeclareMathOperator{\Tr}{Tr}

\newcommand{\relmiddle}[1]{\mathrel{}\middle#1\mathrel{}}
\newcommand{\set}[2]{\left\{ #1 \relmiddle| #2 \right\}}

\newcommand{\software}[1]{\textsc{#1}}
\newcommand{\term}[1]{\emph{#1}}

\begin{document}



\title{
    Algorithm for spin symmetry operation search
}
\shorttitle{
    Algorithm for spin symmetry search
}


\cauthor[a]{Kohei}{Shinohara}{kshinohara0508@gmail.com}{} 
\author[b]{Atsushi}{Togo}{}{} 
\author[d]{Hikaru}{Watanabe}{}{} 
\author[d]{Takuya}{Nomoto} 
\author[a,c,e]{Isao}{Tanaka}{}{} 
\author[d,f]{Ryotaro}{Arita}{}{} 

\aff[a]{Department of Materials Science and Engineering, Kyoto University, \city{Sakyo}, Kyoto 606-8501, \country{Japan}}
\aff[b]{Center for Basic Research on Materials, National Institute for Materials Science, \city{Tsukuba}, Ibaraki 305-0047, \country{Japan}}
\aff[c]{Center for Elements Strategy Initiative for Structural Materials, Kyoto University, \city{Sakyo}, Kyoto 606-8501, \country{Japan}}
\aff[d]{Research Center for Advanced Science and Technology, University of Tokyo, \city{Meguro-ku}, Tokyo 153-8904, \country{Japan}}
\aff[e]{Nanostructures Research Laboratory, Japan Fine Ceramics Center, \city{Nagoya} 456-8587, \country{Japan}}
\aff[f]{RIKEN, Center for Emergent Matter Science, \city{Saitama} 351-0198, \country{Japan}}


\shortauthor{K. Shinohara, A. Togo, H. Watanabe, T. Nomoto, I. Tanaka, and R. Arita}







\maketitle

\begin{synopsis}
This paper presents an algorithm for determining the spin symmetry operations of a given spin arrangement.
Spin symmetry operations of a spin space group act simultaneously on both the spatial and spin coordinates of the spin arrangement.
\end{synopsis}



\begin{abstract}

A spin space group provides a suitable way to fully exploit the symmetry of a spin arrangement with a negligible spin--orbit coupling.
There has been a growing interest in applying spin symmetry analysis with the spin space group in the field of magnetism.
However, there is no established algorithm to search for spin symmetry operations of the spin space group.
This paper presents an exhaustive algorithm for determining spin symmetry operations of commensurate spin arrangements.
The present algorithm searches for spin symmetry operations from the symmetry operations of a corresponding nonmagnetic crystal structure and determines their spin-rotation parts by solving a Procrustes problem.
An implementation is distributed under a permissive free software license in \textsc{spinspg} v0.1.1: \url{https://github.com/spglib/spinspg}.

\end{abstract}





\section{\label{sec:introduction}Introduction}

When the spin--orbit coupling (SOC) is negligible, a spin space group is an appropriate concept to fully exploit the symmetry of a corresponding spin arrangement \cite{LITVIN1974538,Opechowski1986,PhysRevX.12.021016,2105.12738}.
The spin arrangement comprises a crystal structure and magnetic moments.
A spin symmetry operation of the spin space group is assumed to act on the spatial and spin coordinates simultaneously, generalizing a magnetic symmetry operation of a magnetic space group.
The spin space group was first introduced to analyze an extra symmetry of a spin Hamiltonian for neutron scattering experiments \cite{doi:10.1063/1.1708514,doi:10.1098/rspa.1966.0211} and has recently been applied to the field of magnetism \cite{PhysRevX.12.040501}: for example, an analysis of a symmetry-adapted tensor of transport properties with negligible SOC \cite{PhysRevLett.119.187204,Zhang2018}, symmetries of a spin Hamiltonian \cite{PhysRevB.106.144433}, magnon band structure \cite{PhysRevB.105.064430}, and classification of antiferromagnetism \cite{PhysRevX.12.031042}.

When we consider spin space groups of given spin arrangements, we have to first exhaustively search for their spin symmetry operations.
To the best of our knowledge, there is no rigorous algorithm to search for spin symmetry operations.
Therefore, the development of an algorithm and its implementation would benefit the spin symmetry analysis.

There are a few differences between spin space groups and magnetic space groups in terms of symmetry search algorithms.
First, a spin space group may contain nontrivial operations acting on only spin coordinates, called a spin-only group, which complicates the group structure of the spin space group.
Second, although we only need to consider at most double enlarged cells for magnetic space groups, the unit cell size may arbitrarily change between a spin arrangement and its nonmagnetic correspondence.
Lastly, spin rotations, which simultaneously act on magnetic moments, do not have to belong to a crystallographic point group.

Here, we present a rigorous and robust algorithm for determining spin symmetry operations of a given commensurate spin arrangement, extending our magnetic symmetry operation search algorithm \cite{spglibv2}.
The present algorithm fully exploits the group structure of the spin space groups and outputs spin symmetry operations as a coset decomposition of the spin space group, based on the seminal works by Litvin and Opechowski \cite{Litvin:a09793,LITVIN1974538, Litvin:a14103}.
We explicitly denote basis vectors of space groups and spin space groups and employ a lattice algorithm to deal with the case when the unit cell size varies with and without magnetic moments.
We search for the spin-rotation parts from three-dimensional orthogonal groups by solving a well-known optimization problem called a Procrustes problem \cite{10.1093/acprof:oso/9780198510581.001.0001}.
Note that we restrict the present algorithm to commensurate spin arrangements in order to use similar inputs and outputs to an existing space group search implementation \cite{spglibv1}.
The implementation is distributed under the BSD 3-clause license in \textsc{spinspg} v0.1.1 on top of a crystal symmetry search algorithm \cite{spglibv1}.
For magnetic crystal structures tabulated in \textsc{magndata} \cite{Gallego:ks5532}, the present algorithm and implementation have been used to identify physical properties free from SOC \cite{spinspg-application}.

This paper is organized as follows.
In Sec.~\ref{sec:group-structure}, we give definitions of spin space groups and their derived groups.
In Sec.~\ref{sec:symmetry-search}, we provide an algorithm for determining a spin-only group, spin translation group, and spin space group of a given spin arrangement.
In Sec.~\ref{sec:examples}, we demonstrate the present spin symmetry operation search to a spin arrangement of a NiAs-type $\mathrm{CrSe}$.
The notations and terminology in this paper are summarized in Table~\ref{tab:terminology}.





\section{\label{sec:group-structure}Group structure of spin space group}

We provide the definitions of spin symmetry operations and spin arrangements in Sec.~\ref{sec:gs:spin-symmetry-operation}.
We define the spin space group in Sec.~\ref{sec:gs:spin-space-group}.
Then, we introduce its derived groups, the spin-only group (Sec.~\ref{sec:gs:spin-only-group}) and spin translation group (Sec.~\ref{sec:gs:spin-translation-group}).
Although these groups were already discussed in \citeasnoun{Litvin:a09793}, \citeasnoun{LITVIN1974538}, and \citeasnoun{Litvin:a14103}, we consider it beneficial to summarize these results because we fully exploit the group structure of the spin space group in searching for spin symmetry operations.

We note that spin-only groups and spin translation groups complicate the group structure of the spin space groups.
For example, a spin point group, which ignores translation parts of spin symmetry operations of a spin space group, cannot be computed without going through the spin space group due to the existence of a nontrivial spin translation group in general.
Although the analysis of spin point groups is not required to determine spin symmetry operations, we discuss the group structure of spin point groups in Appendix~\ref{appx:spin-point-group} for completeness.

\subsection{\label{sec:gs:spin-symmetry-operation}Spin symmetry operation and spin arrangement}

A \term{spin symmetry operation} comprises a spatial operation in the three-dimensional Euclidean group $\mathrm{E}(3)$ and a spin rotation in the three-dimensional orthogonal group $\mathrm{O}(3)$.
The spin symmetry operation $(g, \bm{W}) \in \mathrm{E}(3) \times \mathrm{O}(3)$ acts on a pair of position $\bm{r}$ and magnetic moments $\bm{m}$ as
\begin{align}
  (g, \bm{W}) (\bm{r}, \bm{m}) = (g \bm{r}, \bm{Wm}).
\end{align}
The product of two spin symmetry operations $(g, \bm{W})$ and $(g', \bm{W}')$ is defined as
\begin{align}
  (g, \bm{W}) (g', \bm{W}') = (gg', \bm{W}\bm{W}').
\end{align}
Although uncommon in crystallography, we suppose both $g$ and $\bm{W}$ are represented with Cartesian coordinates for later convenience.
We denote basis vectors
\begin{align}
  \bm{A}
    = (\bm{a}_{1}, \bm{a}_{2}, \bm{a}_{3})
    = \begin{pmatrix}
      a_{1x} & a_{2x} & a_{3x} \\
      a_{1y} & a_{2y} & a_{3y} \\
      a_{1z} & a_{2z} & a_{3z} \\
    \end{pmatrix}.
\end{align}
When $g$ with a matrix part $\bm{R}$ and a translation part $\bm{v}$ are represented with $\bm{A}$, we explicitly write $g = ( \overline{\bm{R}}, \overline{\bm{v}} )_{\bm{A}}$, where $\bm{R} = \bm{A}\overline{\bm{R}}\bm{A}^{-1}$ and $\bm{v} = \bm{A}\overline{\bm{v}}$.
A \term{spin arrangement} is a set of pairs of a crystal structure and magnetic moments.

\subsection{\label{sec:gs:spin-space-group}Spin space group}

Let $\mathcal{G}$ be a subgroup of $\mathrm{E}(3) \times \mathrm{O}(3)$.
When the following $\mathcal{F}(\mathcal{G})$ and $\mathcal{D}(\mathcal{G})$ are space groups, $\mathcal{G}$ is called a \term{spin space group} \cite{LITVIN1974538},
\begin{align}
  \mathcal{F}(\mathcal{G})
    &= \set{ g \in \mathrm{E}(3) }{ \exists \bm{W} \in \mathrm{O}(3) \,s.t.\, (g, \bm{W}) \in \mathcal{G} } \\
  \mathcal{D}(\mathcal{G})
    &= \set{ g \in \mathrm{E}(3) }{ (g, \bm{E}) \in \mathcal{G} },
\end{align}
where $\bm{E}$ stands for the identity matrix.
For a spin space group $\mathcal{G}$, we call $\mathcal{F}(\mathcal{G})$ a \term{family space group} and $\mathcal{D}(\mathcal{G})$ a \term{maximal space subgroup}.
The maximal space subgroup $\mathcal{D}(\mathcal{G})$ is a normal subgroup of $\mathcal{G}$.
On the contrary, the family space group $\mathcal{F}(\mathcal{G})$ is not a subgroup of $\mathcal{G} \times \mathrm{O}(3)$ in general.
Although \citeasnoun{LITVIN1974538} did not impose the condition that $\mathcal{D}(\mathcal{G})$ is crystallographic, we impose this condition to guarantee that a given spin arrangement is commensurate.
We confine our discussion and algorithms to commensurate spin arrangements.

A spin symmetry operation $(g, \bm{W})$ is transformed by a transformation $(\bm{P}, \bm{p})$ on the spatial coordinates and a transformation matrix $\bm{Q}$ on the spin coordinates as
\begin{align}
  \label{eq:ssg-transformation}
  (g, \bm{W}) \mapsto \left( (\bm{P}, \bm{p})^{-1} g (\bm{P}, \bm{p}), \bm{Q}^{-1} \bm{W} \bm{Q} \right).
\end{align}
Two spin space groups $\mathcal{G}_{1}$ and $\mathcal{G}_{2}$ belong to the same \term{spin-space-group type} if they are transformed to the other by a pair of an orientation-preserving transformation $(\bm{P}, \bm{p})$\footnote{
  A transformation $(\bm{P}, \bm{p})$ is called orientation-preserving if $\det \bm{P} > 0$.
} on the spatial coordinates and a transformation matrix $\bm{Q}$ on the spin coordinates.

In the spin space group, we can consider rotating magnetic moments independently with spatial coordinates.
On the other hand, we consider rotating magnetic moments only in association with spatial rotations and time-reversal operations in the magnetic space group \cite{litvin2014magnetic}.
Thus, the spin space group can be regarded as a supergroup of the magnetic space group under an appropriate correspondence between spin symmetry operations and magnetic symmetry operations as discussed in Appendix~\ref{appx:relationship-with-msg}.

\subsection{\label{sec:gs:spin-only-group}Spin-only group}

A \term{spin-only group} of a spin space group $\mathcal{G}$ is a set of spin symmetry operations of $\mathcal{G}$ with identity spatial operations,
\begin{align}
  \mathcal{P}_{\mathrm{so}}(\mathcal{G})
  =
  \set{ \bm{W} \in \mathrm{O}(3) }{ ((\bm{E}, \bm{0}), \bm{W}) \in \mathcal{G} }.
\end{align}
A direct product of an identity in spatial coordinates and $\mathcal{P}_{\mathrm{so}}(\mathcal{G})$, $1 \times \mathcal{P}_{\mathrm{so}}(\mathcal{G})$, is a subgroup of $\mathcal{G}$\footnote{
  We denote a trivial group, consisting of a single element, as $1$.
}.

\subsection{\label{sec:gs:spin-translation-group}Spin translation group}

A \term{spin translation group} of a spin space group $\mathcal{G}$ is a set of spin symmetry operations with identity rotations in spatial coordinates,
\begin{align}
  \mathcal{G}_{\mathrm{st}}(\mathcal{G})
  =
  \set{ ((\bm{E}, \bm{v}), \bm{W}) }{ ((\bm{E}, \bm{v}), \bm{W}) \in \mathcal{G}}.
\end{align}
\citeasnoun{Litvin:a09793} classified the spin translation groups under the transformation in Eq.~\eqref{eq:ssg-transformation}.

We denote a translation subgroup and a point group of space group $\mathcal{R}$ as $\mathcal{T}(\mathcal{R})$ and $\mathcal{P}(\mathcal{R})$, respectively,
\begin{align}
  \mathcal{T}(\mathcal{R}) &= \set{ (\bm{E}, \bm{t}) }{ (\bm{E}, \bm{t}) \in \mathcal{R} } \\
  \mathcal{P}(\mathcal{R}) &= \set{ \bm{W} }{ \exists \bm{v} \, s.t.\, (\bm{W}, \bm{v}) \in \mathcal{R} }.
\end{align}
Then, the spin-only group $\mathcal{P}_{\mathrm{so}}(\mathcal{G})$ and translation subgroup of $\mathcal{D}(\mathcal{G})$ are a normal subgroup of $\mathcal{G}_{\mathrm{st}}(\mathcal{G})$.
Because $(\mathcal{T}(\mathcal{D}(\mathcal{G})) \times 1) \cap (1 \times \mathcal{P}_{\mathrm{so}}(\mathcal{G})) = 1$, their direct product $\mathcal{T}(\mathcal{D}(\mathcal{G})) \times \mathcal{P}_{\mathrm{so}}(\mathcal{G})$ is also a normal subgroup of $\mathcal{G}_{\mathrm{st}}(\mathcal{G})$.
Thus, we can consider a factor group $\mathcal{G}_{\mathrm{st}}(\mathcal{G}) / ( \mathcal{T}(\mathcal{D}(\mathcal{G})) \times \mathcal{P}_{\mathrm{so}}(\mathcal{G}))$, which is finite for commensurate spin arrangements.
Finally, $\mathcal{G}_{\mathrm{st}}(\mathcal{G})$ is a normal subgroup of $\mathcal{G}$.
The factor group $\mathcal{G} / \mathcal{G}_{\mathrm{st}}(\mathcal{G})$ is isomorphic to $\mathcal{P}(\mathcal{F}(\mathcal{G}))$ and thus finite.




\section{\label{sec:symmetry-search}Spin symmetry operation search}

We provide an algorithm to search for spin symmetry operations from a given spin arrangement represented by the following four objects:
(1) basis vectors of its lattice $\bm{A} = (\bm{a}_{1}, \bm{a}_{2}, \bm{a}_{3})$,
(2) an array of point coordinates of sites in its unit cell $\bm{X} = (\bm{x}_{1}, \cdots, \bm{x}_{N})$,
(3) an array of atomic types of sites in its unit cell $\bm{T} = (t_{1}, \cdots, t_{N})$,
and (4) an array of magnetic moments of sites in its unit cell $\bm{M} = (\bm{m}_{1}, \cdots, \bm{m}_{N})$,
where $N$ is the number of sites in the unit cell.

To search for spin symmetry operations robustly, comparisons of point coordinates and magnetic moments should be performed within tolerances in practice.
We adopt the same absolute tolerance parameter for point coordinates as \citeasnoun{spglibv1}.
For magnetic moments, we use another tolerance parameter $\epsilon_{\mathrm{mag}}$ to identify that two magnetic moments $\bm{W} \bm{m}_{i}$ and $\bm{m}_{\sigma_{g}(i)}$ are equal to the other if
\begin{align}
  \label{eq:magnetic-moments-comparison}
  \norm{ \bm{W} \bm{m}_{i} - \bm{m}_{\sigma_{g}(i)} }_{2} < \epsilon_{\mathrm{mag}}.
\end{align}

The present algorithm extends the authors' previous work on detecting magnetic symmetry operations \cite{spglibv2}.
We first consider a space group of nonmagnetic crystal structure $(\bm{A}, \bm{X}, \bm{T})$ in Sec.~\ref{sec:search:nonmag}.
Next, we search for normal subgroups of the spin space group: the spin-only group $\mathcal{P}_{\mathrm{so}}(\mathcal{G})$ in Sec.~\ref{sec:search:spin-only-group} and the translation subgroup $\mathcal{T}(\mathcal{D}(\mathcal{G}))$ in Sec.~\ref{sec:search:translation-subgroup}.
With these normal subgroups of $\mathcal{G}_{\mathrm{st}}(\mathcal{G})$, we search for coset representatives of the spin translation group $\mathcal{G}_{\mathrm{st}}(\mathcal{G}) / ( \mathcal{T}(\mathcal{D}(\mathcal{G})) \times \mathcal{P}_{\mathrm{so}}(\mathcal{G}))$ in Sec.~\ref{sec:search:spin-translation-group}.
We search for coset representatives of the spin space group $\mathcal{G} / \mathcal{G}_{\mathrm{st}}(\mathcal{G})$ in Sec.~\ref{sec:search:spin-space-group}.
Because the notation in this section is abstract to unambiguously deal with several basis vectors, it may be helpful to read it alongside the examples in Sec.~\ref{sec:examples}.

\subsection{\label{sec:search:nonmag}Space group of nonmagnetic crystal structure}

A candidate for spatial operations of $\mathcal{G}$ can be derived from symmetry operations of a crystal structure $(\bm{A}, \bm{X}, \bm{T})$ ignoring the magnetic moments.
We also employ the space group $\mathcal{S}$ given as a stabilizer of $\mathrm{E}(3)$ preserving $(\bm{A}, \bm{X}, \bm{T})$:
\begin{align}
    \label{eq:space-group-without-magmoms}
    \mathcal{S}
        &= \set{
                g = ( \overline{\bm{R}}, \overline{\bm{v}} )_{\bm{A}} \in \mathrm{E}(3)
            }{
                \begin{array}{l}
                    \exists \sigma_{g} \in \mathfrak{S}_{N}, \forall i = 1, \cdots N, \\
                    \overline{\bm{R}} \bm{x}_{i} + \overline{\bm{v}} \equiv \bm{x}_{\sigma_{g}(i)} \, (\mathrm{mod} \, 1) \\
                    t_{i} = t_{\sigma_{g}(i)}
                \end{array}
            },
\end{align}
where $\mathfrak{S}_{N}$ is a symmetric group of degree $N$.
Note that $g = ( \overline{\bm{R}}, \overline{\bm{v}} )_{\bm{A}}$ maps point coordinates $\bm{x}_{i}$ to $\overline{\bm{R}} \bm{x}_{i} + \overline{\bm{v}}$.
The mapped point coordinates coincide with point coordinates in $\bm{X}$ up to modulo one, inducing permutation $\sigma_{g}$.

The existing crystal symmetry search algorithm \cite{spglibv1} can find primitive basis vectors $\bm{A}_{\mathcal{S}}$ of $\mathcal{T}(\mathcal{S})$ and coset decomposition of $\mathcal{S}$ over $\mathcal{T}_{\bm{A}}$, where we write a translation subgroup formed by $\bm{A}$ as
\begin{align}
  \mathcal{T}_{\bm{A}} = \set{ (\bm{E}, \bm{n})_{\bm{A}} }{ \bm{n} \in \mathbb{Z}^{3} }.
\end{align}
The input basis vector $\bm{A}$ can be represented as an integer linear combination of $\bm{A}_{\mathcal{S}}$.
Thus, an integer matrix $\bm{U} \in \mathbb{Z}^{3 \times 3}$ exists such that $\bm{A} = \bm{A}_{\mathcal{S}} \bm{U}$.
The translation subgroup $\mathcal{T}_{\bm{A}_{\mathcal{S}}}$ is decomposed as
\begin{align}
  \label{eq:translation-nonmag-decomp}
  \mathcal{T}_{\bm{A}_{\mathcal{S}}}
    = \bigsqcup_{ \overline{\bm{t}}^{\mathcal{S}} }
      \left( \bm{E}, \overline{\bm{t}}^{\mathcal{S}} \right)_{ \bm{A}_{\mathcal{S}} }
      \mathcal{T}_{\bm{A}},
\end{align}
where $\overline{\bm{t}}^{\mathcal{S}}$ is a centering vector in a unit cell spanned by $\bm{A}$.
The coset decomposition of $\mathcal{S}$ is written as
\begin{align}
  \label{eq:nonmag-space-group-decomposition}
  \mathcal{S}
    &= \bigsqcup_{ \overline{\bm{R}}^{\mathcal{S}} }
      \left(
        \overline{\bm{R}}^{\mathcal{S}},
        \overline{ \bm{v} }_{ \overline{\bm{R}}^{\mathcal{S}} }
      \right)_{ \bm{A}_{\mathcal{S}} }
      \mathcal{T}_{\bm{A}_{\mathcal{S}}}.
\end{align}
Here, $\overline{ \bm{v} }_{ \overline{\bm{R}}^{\mathcal{S}} }$ is a translation part of a symmetry operation with a matrix part $\overline{\bm{R}}^{\mathcal{S}}$.

For the spin arrangement $(\bm{A}, \bm{X}, \bm{T}, \bm{M})$, the spin space group can be expressed as a stabilizer of $\mathcal{S} \times \mathrm{O}(3)$ that preserves $(\bm{A}, \bm{X}, \bm{T}, \bm{M})$,
\begin{align}
  \label{eq:spin-space-group-as-normalizer}
  \mathcal{G}
    &= \set{
      (g, \bm{W}) \in \mathcal{S} \times \mathrm{O}(3)
    }{
      \bm{W} \bm{m}_{i} = \bm{m}_{\sigma_{g}(i)} \, (i = 1, \dots, N)
    }.
\end{align}
This expression serves as the starting point for the spin symmetry operation search.
For notation simplicity, we denote the maximal space subgroup of $\mathcal{G}$ as $\mathcal{D} = \mathcal{D}(\mathcal{G})$.

\subsection{\label{sec:search:spin-only-group}Spin-only group search}

Because a symmetry operation in the spin-only group of $\mathcal{G}$ does not change the order of point coordinates, the spin-only group of $(\bm{A}, \bm{X}, \bm{T}, \bm{M})$ is expressed as
\begin{align}
  \mathcal{P}_{\mathrm{so}}
    &= \set{
        \bm{W} \in \mathrm{O}(3)
      }{
        \bm{W} \bm{m}_{i} = \bm{m}_{i} \quad (i = 1, \dots, N)
      }.
\end{align}
As shown in Table~\ref{tab:spin_only_group}, when a spin space group $\mathcal{G}$ is a stabilizer of a spin arrangement, spin-only groups are classified into four types up to transformations \cite{LITVIN1974538,PhysRevX.12.021016}: \term{nonmagnetic}, \term{collinear}, \term{coplanar}, and \term{noncoplanar} spin arrangements.

In Sec.~\ref{sec:spin-only-group-search-basic}, we provide an algorithm for detecting $\mathcal{P}_{\mathrm{so}}$ using the eigenvalue decomposition of $\bm{M} \bm{M}^{\top}$.
The spin-only group search should be performed within tolerances for magnetic moments in practice.
In fact, selecting appropriate tolerances is challenging.
In Sec.~\ref{sec:spin-only-group-search-robust}, we propose a robust algorithm, building on the previous section, to alleviate the difficulty.

\subsubsection{\label{sec:spin-only-group-search-basic}Spin-only group search by eigenvalue decomposition}

We consider a moment tensor of $\bm{M} \in \mathbb{R}^{3 \times N}$,
\begin{align}
  \label{eq:magnetic-moments-tensor}
  \bm{M} \bm{M}^{\top} = \sum_{i=1}^{N} \bm{m}_{i} \otimes \bm{m}_{i}.
\end{align}
Because $\bm{M} \bm{M}^{\top}$ is a symmetric semi-definite matrix, we can consider its eigenvalue decomposition,
\begin{align}
  \label{eq:magnetic-moments-eigenvalue-decomposition}
  \bm{M} \bm{M}^{\top} = \sum_{r=1}^{3} \sigma_{r} \hat{\bm{n}}_{r} \otimes \hat{\bm{n}}_{r},
\end{align}
where $\sigma_{1} \geq \sigma_{2} \geq \sigma_{3} \geq 0$ and $\{ \hat{\bm{n}}_{r} \}_{r=1}^{3}$ are orthonormal.

The spin-only group is classified based on the eigenvalues $\{ \sigma_{i} \}_{i=1}^{3}$, as summarized in Table~\ref{tab:spin_only_group}.
In a nonmagnetic spin arrangement, all magnetic moments are zero.
In this case, all eigenvalues are zero, $\sigma_{1} = \sigma_{2} = \sigma_{3} = 0$.
In a collinear spin arrangement, all magnetic moments align parallel or antiparallel to a direction $\hat{\bm{n}}_{\parallel}$.
The eigenvector $\hat{\bm{n}}_{1}$ with the largest eigenvalue should be parallel or antiparallel to $\hat{\bm{n}}_{\parallel}$ and the other eigenvalues $\sigma_{2}$ and $\sigma_{3}$ should be zero.
In a coplanar spin arrangement, all magnetic moments align perpendicular to a direction $\hat{\bm{n}}_{\perp}$.
In this case, the eigenvector $\hat{\bm{n}}_{3}$ with the smallest eigenvalue should be parallel or antiparallel to $\hat{\bm{n}}_{\perp}$ and the smallest eigenvalue $\sigma_{3}$ should be zero.
In other cases, the spin arrangement is noncoplanar.
Note that the site order of magnetic moments $\bm{M}$ does not affect the classification because $\bm{M} \bm{M}^{\top}$ in Eq.~\eqref{eq:magnetic-moments-tensor} remains invariant under a permutation of the site order.

\subsubsection{\label{sec:spin-only-group-search-robust}Numerically robust spin-only group search}

\begin{figure}
  \centering
  \includegraphics[width=0.95\columnwidth]{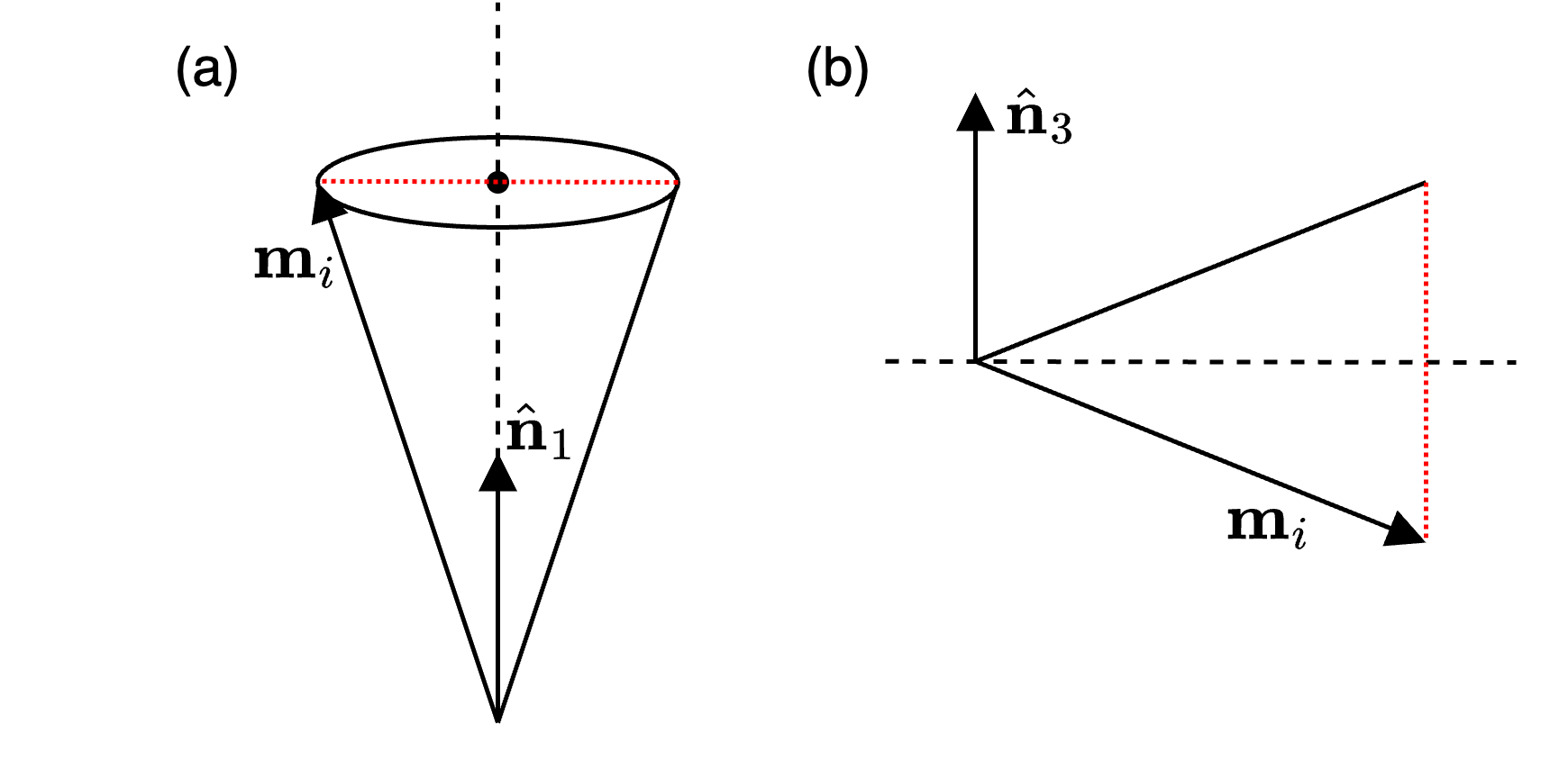}
  \caption{
    Magnetic moments acted by (a) collinear and (b) coplanar spin-only groups.
    (a) A collinear spin-only group along axis $\hat{\bm{n}}_{1}$ is generated from rotations along $\hat{\bm{n}}_{1}$ and mirror operations preserving $\hat{\bm{n}}_{1}$.
    The red dotted line indicates the largest displacement between magnetic moments acted the collinear spin-only group.
    (b) A coplanar spin-only group along axis $\hat{\bm{n}}_{3}$ is generated from a mirror operation perpendicular to $\hat{\bm{n}}_{3}$.
    The red dotted line indicates the largest displacement between magnetic moments acted the coplanar spin-only group.
  }
  \label{fig:spin_only_group_collinear_coplanar}
\end{figure}

The spin-group search algorithm in the previous section requires judging whether the eigenvalues are zero or positive, which needs an additional tolerance parameter.
To reduce the number of tolerance parameters for usability, we modify the spin-group search algorithm solely with the tolerance $\epsilon_{\mathrm{mag}}$ in Eq.~\eqref{eq:magnetic-moments-comparison} as follows.
\begin{enumerate}
  \item We compute the eigenvalues $\sigma_{i}$ and eigenvectors $\hat{\bm{n}}_{i}$ in Eq.~\eqref{eq:magnetic-moments-eigenvalue-decomposition}.
  \item If all magnetic moments are close to zero within $\epsilon_{\mathrm{mag}}$,
    \begin{align}
      \norm{ \bm{m}_{i} }_{2} < \epsilon_{\mathrm{mag}} \quad (i = 1, \dots, N),
    \end{align}
    the spin arrangement is nonmagnetic.
  \item If not, we check if the eigenvector $\hat{\bm{n}}_{1}$ is a parallel or antiparallel direction for all magnetic moments,
    \begin{align}
      2 \norm{ \bm{m}_{i} - (\bm{m}_{i} \cdot \hat{\bm{n}}_{1})\hat{\bm{n}}_{1} }_{2} < \epsilon_{\mathrm{mag}} \quad (\forall i = 1, \dots, N).
    \end{align}
    When spin symmetry operations in a collinear spin-only group prescribed by a direction $\hat{\bm{n}}_{1}$ act on $\bm{m}_{i}$, the acted magnetic moments draw a cone as shown in Fig.~\ref{fig:spin_only_group_collinear_coplanar} (a).
    The left-hand side of the above inequality is the largest displacement between the acted magnetic moments, which corresponds to the diameter of the cone with direction $\hat{\bm{n}}_{1}$.
    If the inequality holds for all magnetic moments, the spin arrangement is collinear.
  \item If not, we check if the eigenvector $\hat{\bm{n}}_{3}$ is a perpendicular direction for all magnetic moments,
    \begin{align}
      2 \left| (\bm{m}_{i} \cdot \hat{\bm{n}}_{3}) \right| < \epsilon_{\mathrm{mag}}
      \quad (\forall i = 1, \dots, N).
    \end{align}
    The left-hand side of the above inequality is the largest displacement between magnetic moments acted by a coplanar spin-only group along $\hat{\bm{n}}_{3}$ as shown in Fig.~\ref{fig:spin_only_group_collinear_coplanar} (b).
    If the inequality holds for all magnetic moments, the spin arrangement is coplanar.
  \item Otherwise, the spin arrangement is noncoplanar.
\end{enumerate}



\subsection{\label{sec:search:translation-subgroup}Translation subgroup search}

\begin{figure}
  \centering
  \includegraphics[width=0.95\columnwidth]{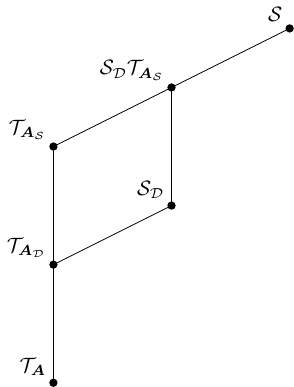}
  \caption{
    Group--subgroup relationship of translation subgroups and space groups derived from spin space group.
    The nodes represent translation subgroups and space groups.
    Each edge indicates that a lower group is a subgroup of an upper group in a diagram.
    Note that $\mathcal{S}_{\mathcal{D}} \mathcal{T}_{\bm{A}_{\mathcal{S}}}$ could be a proper subgroup of $\mathcal{S}$.
  }
  \label{fig:translation_subgroup}
\end{figure}

We search for primitive basis vectors $\bm{A}_{\mathcal{D}}$ of the translation subgroup $\mathcal{T}(\mathcal{D}(\mathcal{G})) = \set{ (\bm{E}, \bm{v}) }{ ((\bm{E}, \bm{v}), \bm{E}) \in \mathcal{G} }$.
The group--subgroup relationships of translation subgroups are shown in Fig.~\ref{fig:translation_subgroup}.
Because $\mathcal{T}_{\bm{A}_{\mathcal{D}}}$ is between $\mathcal{T}_{\bm{A}_{\mathcal{S}}}$ and $\mathcal{T}_{\bm{A}}$, $\mathcal{T}_{\bm{A}} \trianglelefteq \mathcal{T}_{\bm{A}_{\mathcal{D}}} \trianglelefteq \mathcal{T}_{\bm{A}_{\mathcal{S}}}$, we only need to examine finite coset representatives of $\mathcal{T}_{\bm{A}_{\mathcal{S}}} / \mathcal{T}_{\bm{A}}$ for candidates of $\mathcal{T}_{\bm{A}_{\mathcal{S}}} / \mathcal{T}_{\bm{A}_{\mathcal{D}}}$.
For every coset $\left( \bm{E}, \overline{ \bm{t} }^{\mathcal{S}} \right)_{ \bm{A}_{\mathcal{S}} } \mathcal{T}_{\bm{A}}$ in Eq.~\eqref{eq:translation-nonmag-decomp}, we check if $\left( \bm{E}, \overline{ \bm{t} }^{\mathcal{S}} \right)_{ \bm{A}_{\mathcal{S}} }$ preserves magnetic moments $\bm{M}$,
\begin{align}
  \label{eq:translation-subgroup-for-spin}
  \mathcal{T}_{\bm{A}_{\mathcal{D}}}
    &= \set{
        g = \left( \bm{E}, \overline{ \bm{t} }^{\mathcal{S}} \right)_{ \bm{A}_{\mathcal{S}} }
      }{
        \begin{array}{l}
          g \mathcal{T}_{\bm{A}} \in \mathcal{T}_{\bm{A}_{\mathcal{S}}} / \mathcal{T}_{\bm{A}}, \\
          \bm{m}_{i} = \bm{m}_{\sigma_{g}(i)} \, (i = 1, \dots, N)
        \end{array}
      } \mathcal{T}_{\bm{A}}.
\end{align}

An integer matrix $\bm{V} \in \mathbb{Z}^{3 \times 3}$ exists such that $\bm{A}_{\mathcal{D}} = \bm{A}_{\mathcal{S}} \bm{V}$.
A lattice algorithm to generate $\bm{V}$ from centering vectors $\overline{ \bm{t} }^{\mathcal{S}}$ in Eq.~\eqref{eq:translation-subgroup-for-spin} is presented in Appendix~\ref{appx:lattice}.

\subsection{\label{sec:search:spin-translation-group}Spin translation group search}

We can find a spin rotation $\bm{W}$ for a given symmetry operation $g \in \mathcal{S}$ by solving a Procrustes problem \cite{10.1093/acprof:oso/9780198510581.001.0001} as follows.
We write magnetic moments permuted by $g$ as
\begin{align}
  \bm{M}_{g}
    = \left( \bm{m}_{\sigma_{g}(1)}, \cdots, \bm{m}_{\sigma_{g}(N)} \right),
\end{align}
where $\sigma_{g}$ is a permutation of $N$ sites induced by $g$.
Rather than directly searching for $\bm{W} \in \mathrm{O}(3)$ such that $\bm{W} \bm{m}_{i} = \bm{m}_{\sigma_{g}(i)}$, we choose a candidate $\tilde{\bm{W}}$ by solving the following Procrustes problem:
\begin{align}
  \label{eq:procrustes}
  \tilde{\bm{W}}
    &=\argmin_{ \bm{W} \in \mathrm{O}(3) } \norm{ \bm{M}_{g} - \bm{W} \bm{M} }_{F}.
\end{align}
Here the displacement between the magnetic moments and operated ones by $(g, \bm{W})$ is measured by the Frobenius norm $\norm{\cdot}_{F}$.
The solution to Eq.~\eqref{eq:procrustes} can be explicitly written as
\begin{align}
  \tilde{\bm{W}} = \bm{Y} \bm{Z}^{\top},
\end{align}
where $\bm{Y}$ and $\bm{Z}$ are orthogonal matrices of the singular value decomposition of $\bm{M}_{g} \bm{M}^{\top}$,
\begin{align}
  \bm{M}_{g} \bm{M}^{\top}
    = \bm{Y} \bm{\Sigma} \bm{Z}^{\top}.
\end{align}
After we obtain $\tilde{\bm{W}}$, $(g, \tilde{\bm{W}})$ is taken as a spin symmetry operation if the condition
\begin{align}
  \label{eq:magmom-check}
  \norm{ \bm{m}_{i} - \tilde{\bm{W}} \bm{m}_{\sigma_{g}(i)} }_{2} < \epsilon_{\mathrm{mag}}
\end{align}
holds for every site $i$.
If Eq.~\eqref{eq:magmom-check} does not hold for some site $i$, we reject $(g, \tilde{\bm{W}})$ as a spin symmetry operation.

For coset representatives of $(\bm{E}, \overline{\bm{t}}^{\mathcal{D}} )_{\bm{A}_{\mathcal{D}}} \mathcal{T}_{\bm{A}_{\mathcal{D}}} \in \mathcal{T}_{\bm{A}_{\mathcal{S}}} / \mathcal{T}_{\bm{A}_{\mathcal{D}}}$, we search for a corresponding spin rotation $\bm{W}_{ \overline{\bm{t}}^{\mathcal{D}} }$ using the above algorithm if it exists.
The coset decomposition of $\mathcal{G}_{\mathrm{st}}$ can be written as
\begin{align}
  \mathcal{G}_{\mathrm{st}}
    &= \set{
      g = \left( (\bm{E}, \overline{\bm{t}}^{\mathcal{D}})_{\bm{A}_{\mathcal{D}}}, \bm{W}_{ \overline{\bm{t}}^{\mathcal{D}} } \right)
    }{
        \begin{array}{l}
          (\bm{E}, \overline{\bm{t}}^{\mathcal{D}})_{\bm{A}_{\mathcal{D}}} \mathcal{T}_{\bm{A}_{\mathcal{D}}} \in \mathcal{T}_{\bm{A}_{\mathcal{S}}} / \mathcal{T}_{\bm{A}_{\mathcal{D}}} \\
          \bm{W}_{ \overline{\bm{t}}^{\mathcal{D}} } \bm{m}_{i} = \bm{m}_{\sigma_{g}(i)} \, (\forall i = 1, \cdots, N)
        \end{array}
    }
  \left(
    \mathcal{T}_{\bm{A}_{\mathcal{D}}} \times \mathcal{P}_{\mathrm{so}}
  \right),
\end{align}
where $\overline{\bm{t}}^{\mathcal{D}}$ takes a distinct translation up to $\mathcal{T}_{\bm{A}_{\mathcal{D}}}$ and we choose $\bm{W}_{\overline{\bm{t}}^{\mathcal{D}} = \bm{0}} = \bm{E}$.

\subsection{\label{sec:search:spin-space-group}Spin space group search}

Because the translation subgroup of $\mathcal{F}(\mathcal{G})$ is $\mathcal{T}_{\bm{A}_{\mathcal{D}}}$, spatial operation parts of $\mathcal{G}$ should belong to a maximal subgroup $\mathcal{S}_{\mathcal{D}}$ of $\mathcal{S}$ with its translation subgroup $\mathcal{T}_{\bm{A}_{\mathcal{D}}}$.
Figure~\ref{fig:translation_subgroup} shows the group--subgroup relationship between $\mathcal{S}$ and $\mathcal{S}_{\mathcal{D}}$.
A rotation $\bm{A}_{\mathcal{S}} \overline{\bm{R}}^{\mathcal{S}} \bm{A}_{\mathcal{S}}^{-1} \in \mathcal{P}(\mathcal{S})$ is compatible with $\mathcal{T}_{\bm{A}_{\mathcal{D}}}$ if $\bm{V}^{-1} \overline{\bm{R}}^{\mathcal{S}} \bm{V}$ is an integer matrix \cite{PhysRevB.77.224115}.
Thus, the compatible subgroup $\mathcal{S}_{\mathcal{D}}$ is written as
\begin{align}
  \label{eq:space-subgroup-with-spin-unit-cell}
  \mathcal{S}_{\mathcal{D}}
    &= \set{
        \left(
          \overline{\bm{R}}^{\mathcal{S}}, \overline{ \bm{v} }_{ \overline{\bm{R}}^{\mathcal{S}} } + \overline{\bm{t}}_{\overline{\bm{R}}^{\mathcal{S}}}
        \right)_{ \bm{A}_{\mathcal{S}} }
    }{
      \begin{array}{l}
        \left( \overline{\bm{R}}^{\mathcal{S}}, \overline{ \bm{v} }_{ \overline{\bm{R}}^{\mathcal{S}} } \right) \mathcal{T}_{\bm{A}_{\mathcal{S}}} \in \mathcal{S} / \mathcal{T}_{\bm{A}_{\mathcal{S}}} \\
        \bm{V}^{-1} \overline{\bm{R}}^{\mathcal{S}} \bm{V} \in \mathbb{Z}^{3 \times 3} \\
        \left( \bm{E}, \overline{\bm{t}}_{\overline{\bm{R}}^{\mathcal{S}}} \right)_{\bm{A}_{\mathcal{S}}} \mathcal{T}_{\bm{A}} \in \mathcal{T}_{\bm{A}_{\mathcal{S}}} / \mathcal{T}_{\bm{A}}
      \end{array}
    }
    \mathcal{T}_{\bm{A}_{\mathcal{D}}},
\end{align}
where $\overline{\bm{t}}_{\overline{\bm{R}}^{\mathcal{S}}}$ is a centering vector in $\mathcal{T}_{\bm{A}_{\mathcal{S}}} / \mathcal{T}_{\bm{A}}$.
The additional centering vector $\overline{\bm{t}}_{\overline{\bm{R}}^{\mathcal{S}}}$ is necessary for $\mathcal{S}_{\mathcal{D}}$ to form a group \cite{nebe20061,Stokes:vk5013}.

For a coset representative of $\mathcal{S}_{\mathcal{D}} / \mathcal{T}_{\bm{A}_{\mathcal{D}}}$, a corresponding spin-rotation part $\bm{W}_{\overline{\bm{R}}^{\mathcal{S}}}$ can also be determined by solving the Procrustes problem as presented in Sec.~\ref{sec:search:spin-translation-group}.
Finally, we obtain all spin symmetry operations in a coset decomposition
\begin{align}
  \mathcal{G}
    = \set{
        \left(
          \left(
            \overline{\bm{R}}^{\mathcal{S}},
            \overline{ \bm{v} }_{ \overline{\bm{R}}^{\mathcal{S}} } + \overline{\bm{t}}_{\overline{\bm{R}}^{\mathcal{S}}}
          \right)_{ \bm{A}_{\mathcal{S}} },
          \bm{W}_{\overline{\bm{R}}^{\mathcal{S}}}
        \right)
    }{
      \begin{array}{l}
        \left(
          \overline{\bm{R}}^{\mathcal{S}},
          \overline{ \bm{v} }_{ \overline{\bm{R}}^{\mathcal{S}} } + \overline{\bm{t}}_{\overline{\bm{R}}^{\mathcal{S}}}
        \right)_{ \bm{A}_{\mathcal{S}} }
        \mathcal{T}_{\bm{A}_{\mathcal{D}}} \in \mathcal{S}_{\mathcal{D}} / \mathcal{T}_{\bm{A}_{\mathcal{D}}} \\
        \bm{W}_{ \overline{\bm{R}}^{\mathcal{S}} } \bm{m}_{i} = \bm{m}_{\sigma_{g}(i)} \, (\forall i = 1, \cdots, N)
      \end{array}
    } \mathcal{G}_{\mathrm{st}}.
\end{align}



\section{\label{sec:examples}Examples of spin symmetry operation search}

\begin{figure}
  \centering
  \includegraphics[width=0.95\columnwidth]{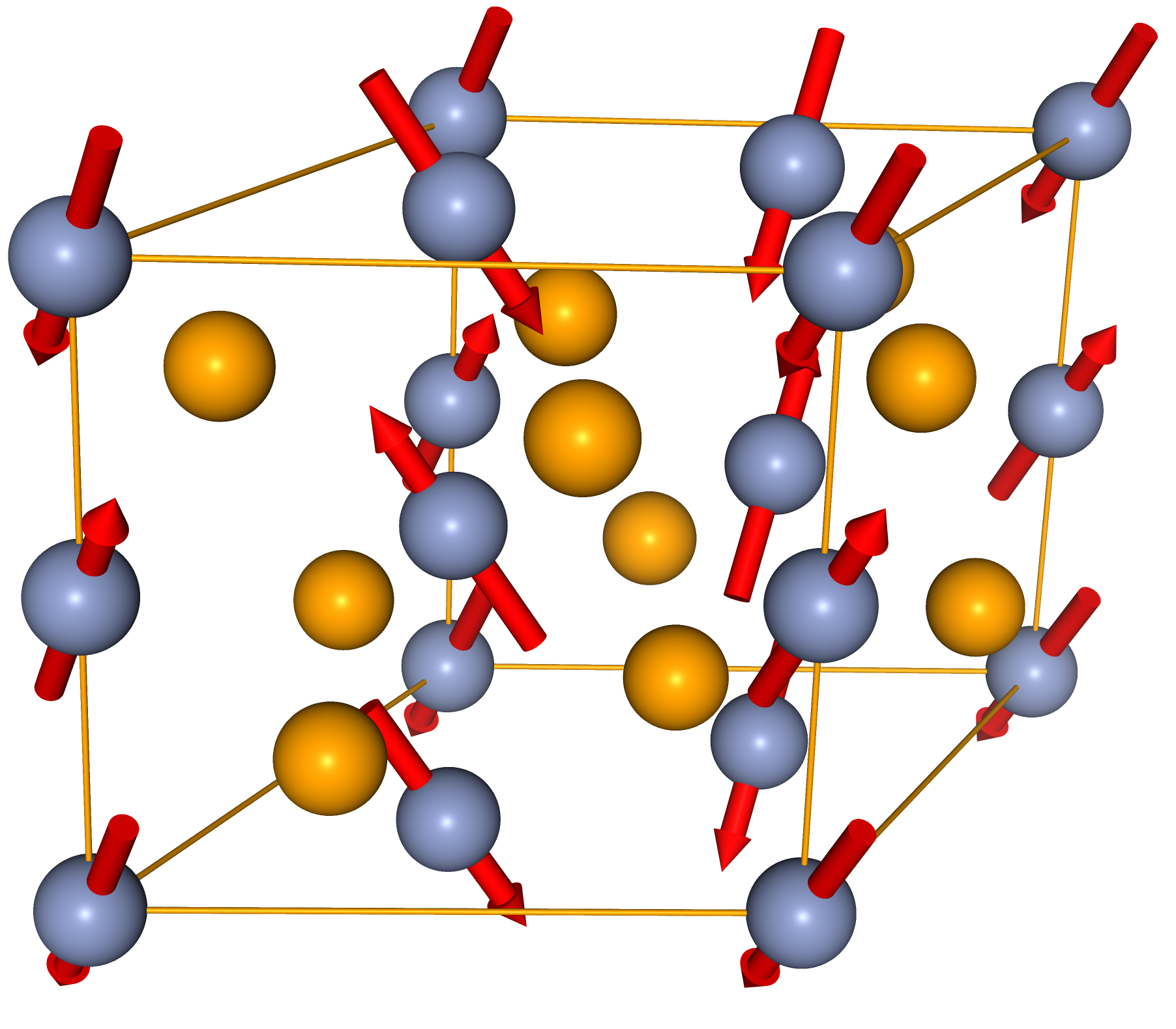}
  \caption{
    Spin arrangement example for NiAs-type CrSe.
    The gray and orange balls represent Cr and Se atoms, respectively.
    The red arrows denote magnetic moments of Cr atoms with equal magnitudes.
  }
  \label{fig:example_spin_arrangement}
\end{figure}

We consider a spin arrangement of the NiAs-type CrSe \cite{PhysRev.122.1402,Litvin:a09793}  (\#2.35 of \software{magndata} \cite{Gallego:ks5532}) as an example of spin symmetry operation search, illustrated in Fig.~\ref{fig:example_spin_arrangement}.
The hexagonal lattice of CrSe has the following basis vectors
\begin{align*}
  \bm{A} = \begin{pmatrix}
    a & -\frac{1}{2} a       & 0 \\
    0 & \frac{\sqrt{3}}{2} a & 0 \\
    0 & 0                    & c \\
  \end{pmatrix}.
\end{align*}
The fractional coordinates and magnetic moments of atoms are as follows.
\begin{align*}
  \mathrm{Cr}:
    &\bm{x}_{1} = \left( 0, 0, 0 \right)^{\top},
    \bm{m}_{1} = \left( -\frac{1}{2} m_{x}, -\frac{\sqrt{3}}{2} m_{x}, -m_{z} \right)^{\top} \\
  \mathrm{Cr}:
    &\bm{x}_{2} = \left( \frac{1}{3}, \frac{2}{3}, 0 \right)^{\top},
    \bm{m}_{2} = \left( m_{x}, 0, -m_{z} \right)^{\top} \\
  \mathrm{Cr}:
    &\bm{x}_{3} = \left( \frac{2}{3}, \frac{1}{3}, 0 \right)^{\top},
    \bm{m}_{3} = \left( -\frac{1}{2} m_{x}, \frac{\sqrt{3}}{2} m_{x}, -m_{z} \right)^{\top} \\
  \mathrm{Cr}:
    &\bm{x}_{4} = \left( 0, 0, \frac{1}{2} \right)^{\top},
    \bm{m}_{4} = \left( \frac{1}{2} m_{x}, \frac{\sqrt{3}}{2} m_{x}, m_{z} \right)^{\top} \\
  \mathrm{Cr}:
    &\bm{x}_{5} = \left( \frac{1}{3}, \frac{2}{3}, \frac{1}{2} \right)^{\top},
    \bm{m}_{5} = \left( -m_{x}, 0, m_{z} \right)^{\top} \\
  \mathrm{Cr}:
    &\bm{x}_{6} = \left( \frac{2}{3}, \frac{1}{3}, \frac{1}{2} \right)^{\top},
    \bm{m}_{6} = \left( \frac{1}{2} m_{x}, -\frac{\sqrt{3}}{2} m_{x}, m_{z} \right)^{\top} \\
  \mathrm{Se}:
    &\bm{x}_{7} = \left( 0, \frac{1}{3}, \frac{1}{4} \right)^{\top},
    \bm{m}_{7} = \left( 0, 0, 0 \right)^{\top} \\
  \mathrm{Se}:
    &\bm{x}_{8} = \left( \frac{1}{3}, 0, \frac{1}{4} \right)^{\top},
    \bm{m}_{8} = \left( 0, 0, 0 \right)^{\top} \\
  \mathrm{Se}:
    &\bm{x}_{9} = \left( \frac{1}{3}, \frac{1}{3}, \frac{3}{4} \right)^{\top},
    \bm{m}_{9} = \left( 0, 0, 0 \right)^{\top} \\
  \mathrm{Se}:
    &\bm{x}_{10} = \left( 0, \frac{2}{3}, \frac{3}{4} \right)^{\top},
    \bm{m}_{10} = \left( 0, 0, 0 \right)^{\top} \\
  \mathrm{Se}:
    &\bm{x}_{11} = \left( \frac{2}{3}, 0, \frac{3}{4} \right)^{\top},
    \bm{m}_{11} = \left( 0, 0, 0 \right)^{\top} \\
  \mathrm{Se}:
    &\bm{x}_{12} = \left( \frac{2}{3}, \frac{2}{3}, \frac{1}{4} \right)^{\top},
    \bm{m}_{12} = \left( 0, 0, 0 \right)^{\top}
\end{align*}
Here, the magnetic moments are represented with Cartesian coordinates.

\subsection{Space group of nonmagnetic crystal structure}

The space-group type of $\mathcal{S}$ for the crystal structure ignoring magnetic moments is $P6_{3}/mmc$ (No. 194).
One of the primitive basis vectors for $\mathcal{S}$ and the transformation matrix are
\begin{align*}
  \bm{A}_{\mathcal{S}}
    &= \begin{pmatrix}
        0 & -\frac{1}{2} a & 0 \\
        -\frac{\sqrt{3}}{3} a & \frac{\sqrt{3}}{6} a & 0 \\
        0 & 0 & -c \\
    \end{pmatrix} \\
  \bm{U}
    &= \begin{pmatrix}
      -1 & -1 & 0 \\
      -2 & 1 & 0 \\
      0 & 0 & -1 \\
    \end{pmatrix}
\end{align*}
with $\bm{A} = \bm{A}_{\mathcal{S}} \bm{U}$ as defined in Sec.~\ref{sec:search:nonmag}.
There are three coset representatives for $\mathcal{T}_{\bm{A}_{\mathcal{S}}} / \mathcal{T}_{\bm{A}}$,
\begin{align}
  \label{eq:example-centerings}
  \mathcal{T}_{\bm{A}_{\mathcal{S}}}
    &= g_{1} \mathcal{T}_{\bm{A}}
      \sqcup g_{2} \mathcal{T}_{\bm{A}}
      \sqcup g_{3} \mathcal{T}_{\bm{A}} \\
    g_{1} &= \left( \bm{E}, (0, 0, 0)^{\top} \right)_{\bm{A}_{\mathcal{S}}} \nonumber \\
    g_{2} &= \left( \bm{E}, (-1, -1, 0)^{\top} \right)_{\bm{A}_{\mathcal{S}}} \nonumber \\
    g_{3} &= \left( \bm{E}, (-1, 0, 0)^{\top} \right)_{\bm{A}_{\mathcal{S}}}. \nonumber
\end{align}

\subsection{Spin-only group}

The moment tensor of $\bm{M}$ is given by
\begin{align*}
  \bm{M} \bm{M}^{\top}
    &= \begin{pmatrix}
      3 m_{x}^{2} & 0 & 0 \\
      0 & 3 m_{x}^{2} & 0 \\
      0 & 0 & 6 m_{z}^{2} \\
    \end{pmatrix}.
\end{align*}
Its eigenvalues $3 m_{x}^{2}$, $3 m_{x}^{2}$, and $6 m_{z}^{2}$ are all positive.
Consequently, the spin arrangement is noncoplanar with $\mathcal{P}_{\mathrm{so}} = 1$.

\subsection{Translation subgroup of maximal space subgroup}

The translation $g_{1}$ in Eq.~\eqref{eq:example-centerings} belongs to $\mathcal{T}_{\mathcal{A}_{\mathcal{D}}}$.
On the other hand, $g_{2}$ does not belong to $\mathcal{T}_{\mathcal{A}_{\mathcal{D}}}$ because it maps $\bm{x}_{1}$ to $\bm{x}_{3}$, whereas $\bm{m}_{1} \neq \bm{m}_{3}$.
Similarly, $g_{3}$ does not belong to $\mathcal{T}_{\mathcal{A}_{\mathcal{D}}}$.
Therefore, $\mathcal{T}_{\bm{A}_{\mathcal{D}}}$ is identical to $\mathcal{T}_{\bm{A}_{\mathcal{S}}}$ with $\bm{A}_{\mathcal{D}} = \bm{A}_{\mathcal{S}} \bm{V}$ as defined in Sec.~\ref{sec:search:translation-subgroup} and we can choose $\bm{V} = \bm{E}$.

\subsection{\label{sec:example:gst}Coset representatives of spin translation group}

There are three candidates for the spatial operation parts of coset representatives of $\mathcal{G}_{\mathrm{st}}$: $g_{1}$, $g_{2}$, and $g_{3}$.
For translation $g_{1}$ in Eq.~\eqref{eq:example-centerings}, we obtain
\begin{align*}
  \bm{M}_{g_{1}} \bm{M}^{\top}
    = \begin{pmatrix}
      3 m_{x}^{2} & 0 & 0 \\
      0 & 3 m_{x}^{2} & 0 \\
      0 & 0 & 6 m_{z}^{2} \\
    \end{pmatrix}.
\end{align*}
From the singular value decomposition of $\bm{M}_{g_{1}} \bm{M}^{\top}$, we calculate the spin-rotation part $\tilde{\bm{W}}_{1}$ for $g_{1}$ as $\tilde{\bm{W}}_{1} = \bm{E}$.
For translation $g_{2}$, the singular value decomposition of $\bm{M}_{g_{2}} \bm{M}^{\top}$ is
\begin{align*}
  \bm{M}_{g_{2}} \bm{M}^{\top}
    = \begin{pmatrix}
      -\frac{1}{2} & \frac{\sqrt{3}}{2} & 0 \\
      -\frac{\sqrt{3}}{2} & -\frac{1}{2} & 0 \\
      0 & 0 & 1 \\
    \end{pmatrix}
    \begin{pmatrix}
      3 m_{x}^{2} & 0 & 0 \\
      0 & 3 m_{x}^{2} & 0 \\
      0 & 0 & 6 m_{x}^{2} \\
    \end{pmatrix}
    \begin{pmatrix}
      1 & 0 & 0 \\
      0 & 1 & 0 \\
      0 & 0 & 1 \\
    \end{pmatrix}^{\top}.
\end{align*}
Hence, a candidate for a spin-rotation part with $g_{2}$ is
\begin{align*}
  \tilde{\bm{W}}_{2}
    = \begin{pmatrix}
      -\frac{1}{2} & \frac{\sqrt{3}}{2} & 0 \\
      -\frac{\sqrt{3}}{2} & -\frac{1}{2} & 0 \\
      0 & 0 & 1 \\
    \end{pmatrix}
    \begin{pmatrix}
      1 & 0 & 0 \\
      0 & 1 & 0 \\
      0 & 0 & 1 \\
    \end{pmatrix}^{\top}
    = \begin{pmatrix}
        -\frac{1}{2} & \frac{\sqrt{3}}{2} & 0 \\
        -\frac{\sqrt{3}}{2} & -\frac{1}{2} & 0 \\
        0 & 0 & 1 \\
    \end{pmatrix}.
\end{align*}
Similarly, a candidate for a spin-rotation part with $g_{3}$ is
\begin{align*}
  \tilde{\bm{W}}_{3}
    = \begin{pmatrix}
        -\frac{1}{2} & -\frac{\sqrt{3}}{2} & 0 \\
        \frac{\sqrt{3}}{2} & -\frac{1}{2} & 0 \\
        0 & 0 & 1 \\
    \end{pmatrix}.
\end{align*}
The spin symmetry operations $(g_{i}, \tilde{\bm{W}}_{i}) \, (i=1, 2, 3)$ all preserve magnetic moments.
Consequently, the spin translation group is obtained as
\begin{align*}
  \mathcal{G}_{\mathrm{st}}
    = (g_{1}, \tilde{\bm{W}}_{1}) (\mathcal{T}_{\bm{A}_{\mathcal{D}}} \times 1)
      \sqcup (g_{2}, \tilde{\bm{W}}_{2}) (\mathcal{T}_{\bm{A}_{\mathcal{D}}} \times 1)
      \sqcup (g_{3}, \tilde{\bm{W}}_{3}) (\mathcal{T}_{\bm{A}_{\mathcal{D}}} \times 1).
\end{align*}

\subsection{Coset representatives of spin space group}

Because $\mathcal{T}_{\bm{A}_{\mathcal{D}}}$ and $\mathcal{T}_{\bm{A}_{\mathcal{S}}}$ coincide in this example, $\mathcal{S}$ is fully compatible with $\mathcal{T}_{\bm{A}_{\mathcal{D}}}$, $\mathcal{S}_{\mathcal{D}} = \mathcal{S}$.
For 24 coset representatives of $\mathcal{S}_{\mathcal{D}} / \mathcal{T}_{\bm{A}_{\mathcal{D}}}$, we search for their spin-rotation parts by solving the Procrustes problems.
For example, a six-fold spatial rotoinversion
\begin{align*}
  g = \left(
    \begin{pmatrix}
      -1 & 1 & 0 \\
      -1 & 0 & 0 \\
      0 & 0 & -1 \\
    \end{pmatrix},
    \bm{0}
  \right)_{\bm{A}_{\mathcal{D}}}
  \in \mathcal{S}_{\mathcal{D}}
\end{align*}
gives a permutation of sites $\sigma_{g} = \left( 4, 5, 6, 1, 2, 3, 12, 7, 11, 9, 10, 8 \right)$ and the singular value decomposition of $\bm{M}_{g} \bm{M}$
\begin{align*}
  \bm{M}_{g} \bm{M}
    = \begin{pmatrix}
      -1 & 0 & 0 \\
      0 & -1 & 0 \\
      0 & 0 & -1 \\
    \end{pmatrix}
    \begin{pmatrix}
      3 m_{x}^{2} & 0 & 0 \\
      0 & 3 m_{x}^{2} & 0 \\
      0 & 0 & 6 m_{x}^{2} \\
    \end{pmatrix}
    \begin{pmatrix}
      1 & 0 & 0 \\
      0 & 1 & 0 \\
      0 & 0 & 1 \\
    \end{pmatrix}^{\top},
\end{align*}
which results in $\tilde{\bm{W}} = -\bm{E}$.
Similarly, all of the 24 coset representatives of $\mathcal{S}_{\mathcal{D}} / \mathcal{T}_{\bm{A}_{\mathcal{D}}}$ give spin symmetry operations preserving $\bm{M}$.

A magnetic space group $\mathcal{M}$ of the NiAs-type CrSe is $P31m'$ (BNS number 157.55), which has $|\mathcal{M} / \mathcal{T}_{\bm{A}_{\mathcal{D}}}| = 6$ coset representatives.
Because the spin translation group $\mathcal{G}_{\mathrm{st}}$ contains three-fold spin-rotation parts, these spin symmetry operations ($(g_{2}, \tilde{\bm{W}}_{2})$ and $(g_{3}, \tilde{\bm{W}}_{3})$ in Sec.~\ref{sec:example:gst}) are not captured in $\mathcal{M}$.
Similarly, the spin symmetry operations with six-fold spatial rotations or rotoinversions are not preserved in $\mathcal{M}$.




\section{\label{sec:conclusion}Conclusion}

We have presented the algorithm for determining spin symmetry operations of a given spin arrangement.
The spin-only group is robustly determined from the eigenvalue decomposition of the moment tensor of magnetic moments, $\bm{M}\bm{M}^{\top}$.
We have explicitly considered the three translation subgroups to address the enlargement of the unit cell due to the spin translation group: the translation subgroup spanned by the input basis vectors $\mathcal{T}_{\bm{A}}$, one by the primitive basis vector $\mathcal{T}_{\bm{A}_{\mathcal{S}}}$, and one by the primitive basis vectors for the spin space group $\mathcal{T}_{\bm{A}_{\mathcal{D}}}$.
Spin-rotation parts of the coset representatives of the spin translation group and spin space group are found by solving the Procrustes problem to match the original magnetic moments and permuted ones.
The present algorithm is implemented in \textsc{spinspg} under a permissive license.
In future work, it will be beneficial to identify the spin-space-group type and a suitable transformation from the spin symmetry operations.
Our presented algorithm and implementation will advance spin symmetry analysis in crystallography and condensed matter physics.

\paragraph*{Note added}

After this work was completed, we became aware of recent preprints on the classification and enumeration of spin space-group types \cite{2307.10364,2307.10369,2307.10371}.
Two of these works, \citeasnoun{2307.10364} and \citeasnoun{2307.10371}, appear to use a workflow similar to ours to determine spin symmetry operations although they only provide a brief description and their implementations are not available to the public.

Our presented algorithm uses the Procrustes problem to determine a spin-rotation part $\bm{W}$.
On the other hand, \citeasnoun{2307.10364} directly determines $\bm{W}$ for a symmetry operation $g$ such that selected three magnetic moments $\bm{m}_{i} \quad (i=1, 2, 3)$ are transformed into $\bm{m}_{\sigma_{g}(i)}$, which may not be robust against numerical noises and experimental uncertainty of magnetic moments.

\citeasnoun{2307.10371} identifies spin symmetry operations on top of \textsc{spglib} \cite{spglibv1} similar to ours.
However, they do not incorporate translation subgroups $\mathcal{T}_{\bm{A}_{\mathcal{D}}}$ and $\mathcal{T}_{\bm{A}}$.
This omission makes their algorithm complicated and non-exhaustive.




\appendix


\section{\label{appx:relationship-with-msg}Correspondence between spin symmetry operation and magnetic symmetry operation}

When we consider a spin symmetry operation $((\bm{R}, \bm{v}), \bm{W})$ for a Hamiltonian without the spin--orbit coupling (SOC), we derive the condition that $((\bm{R}, \bm{v}), \bm{W})$ has a corresponding magnetic symmetry operation for the Hamiltonian with SOC.
The SOC introduces an additional term to the Hamiltonian proportional to $\hat{\bm{L}} \cdot \hat{\bm{\sigma}}$, with the angular momentum operator $\hat{\bm{L}}$ and the Pauli matrices $\hat{\bm{\sigma}} = (\hat{\sigma}_{x}, \hat{\sigma}_{y}, \hat{\sigma}_{z})$.
A spin symmetry operation $((\bm{R}, \bm{v}), \bm{W})$ acts on $\hat{\bm{L}}$ and $\hat{\bm{\sigma}}$ as
\begin{align}
  \label{eq:angular-momentum-transformation}
  \hat{\bm{L}}
    &\mapsto (\det \bm{R}) (\det \bm{W}) \left( \sum_{\nu=x,y,z} R_{\mu \nu} \hat{L}_{\nu} \right)_{\mu=x,y,z} \\
  \hat{\bm{\sigma}}
    &\mapsto \left( \sum_{\nu=x,y,z} W_{\mu \nu} \hat{\sigma}_{\nu} \right)_{\mu=x,y,z},
\end{align}
where $\bm{R}$ and $\bm{W}$ are represented with Cartesian coordinates.
Note that $\det \bm{W}$ in Eq.~\eqref{eq:angular-momentum-transformation} reflects a time reversal operation.
Thus, $((\bm{R}, \bm{v}), \bm{W})$ acts on the SOC term as
\begin{align}
  \hat{\bm{L}} \cdot \hat{\bm{\sigma}}
    &\mapsto (\det \bm{R}) (\det \bm{W}) \sum_{\mu, \nu = x, y, z} \left[ \bm{R}^{\top} \bm{W} \right]_{\mu\nu} \hat{L}_{\mu} \hat{\sigma}_{\nu}.
\end{align}
To preserve the SOC term, $(\det \bm{R}) (\det \bm{W}) \bm{R}^{\top} \bm{W}$ should be identity.
Because $(\det \bm{R}) \bm{R}$ and $(\det \bm{W}) \bm{W}$ belong to $\mathrm{SO}(3)$, this condition is equivalent to
\begin{align}
  \label{eq:spin-magnetic-correspondence}
  (\det \bm{R}) \bm{R} = (\det \bm{W}) \bm{W}.
\end{align}
Therefore, if $((\bm{R}, \bm{v}), \bm{W})$ satisfies Eq.~\eqref{eq:spin-magnetic-correspondence}, it has a corresponding magnetic symmetry operation $((\bm{R}, \bm{v}), \det \bm{W})$ with symmetry operation part $(\bm{R}, \bm{v})$ and time-reversal part $\det \bm{W}$.
We identify $\det \bm{W} = 1$ with an identity operation and $\det \bm{W} = -1$ with a time-reversal operation.
Here, we assume that a magnetic symmetry operation $((\bm{R}, \bm{v}), \theta)$ acts on a magnetic moments $\bm{m}$ as $\theta (\det \bm{R})\bm{R} \bm{m}$.

Conversely, a magnetic symmetry operation $((\bm{R}, \bm{v}), \theta)$ is mapped to a spin symmetry operation $((\bm{R}, \bm{v}), \theta \bm{R})$.
We can confirm this mapping satisfies Eq.~\eqref{eq:spin-magnetic-correspondence} by $(\det \theta \bm{R}) \theta \bm{R} = (\det \bm{R}) \bm{R}$.
Because this mapping is injective, the index of a magnetic space group $\mathcal{M}$ in its translation subgroup $\mathcal{T}_{\bm{A}_{\mathcal{D}}}$ is a divisor of the index of a corresponding spin space group $\mathcal{G}$ in $\mathcal{T}_{\bm{A}_{\mathcal{D}}}$,
$|\mathcal{G} / \mathcal{T}_{\bm{A}_{\mathcal{D}}}| \equiv 0 \, (\mathrm{mod} \, |\mathcal{M} / \mathcal{T}_{\bm{A}_{\mathcal{D}}}|)$.



\section{\label{appx:spin-point-group}Spin point group}

We define a \term{family spin point group} of a spin space group $\mathcal{G}$ as
\begin{align}
  \mathcal{B}(\mathcal{G})
    &= \set{
        \bm{W} \in \mathrm{O}(3)
    }{
        \exists g \, s.t. \, (g, \bm{W}) \in \mathcal{G}
    }.
\end{align}
When we write a coset decomposition of $\mathcal{G}$ by its spin translation group $\mathcal{G}_{\mathrm{st}}(\mathcal{G})$ as
\begin{align}
    \mathcal{G}
        = \bigsqcup_{ \bm{R} } \left( (\bm{R}, \bm{v}_{\bm{R}}), \bm{W}_{\bm{R}} \right) \mathcal{G}_{\mathrm{st}}(\mathcal{G}),
\end{align}
A \term{spin point group} of $\mathcal{G}$ is
\begin{align}
    \mathcal{U}(\mathcal{G})
        = \left\{
            \left( \bm{R}, \bm{W}_{\bm{R}} \mathcal{B}(\mathcal{G}_{\mathrm{st}}(\mathcal{G})) \right)
        \right\}_{ \bm{R} \in \mathcal{P}(\mathcal{F}(\mathcal{G})) }.
\end{align}
The spin point group $\mathcal{U}(\mathcal{G})$ is a subgroup of $\mathrm{O}(3) \times (\mathrm{O}(3) / \mathcal{B}(\mathcal{G}_{\mathrm{st}}(\mathcal{G})) )$ and is isomorphic to $\mathcal{G} / \mathcal{G}_{\mathrm{st}}(\mathcal{G})$.

\begin{figure}
  \centering
  \includegraphics[width=0.95\columnwidth]{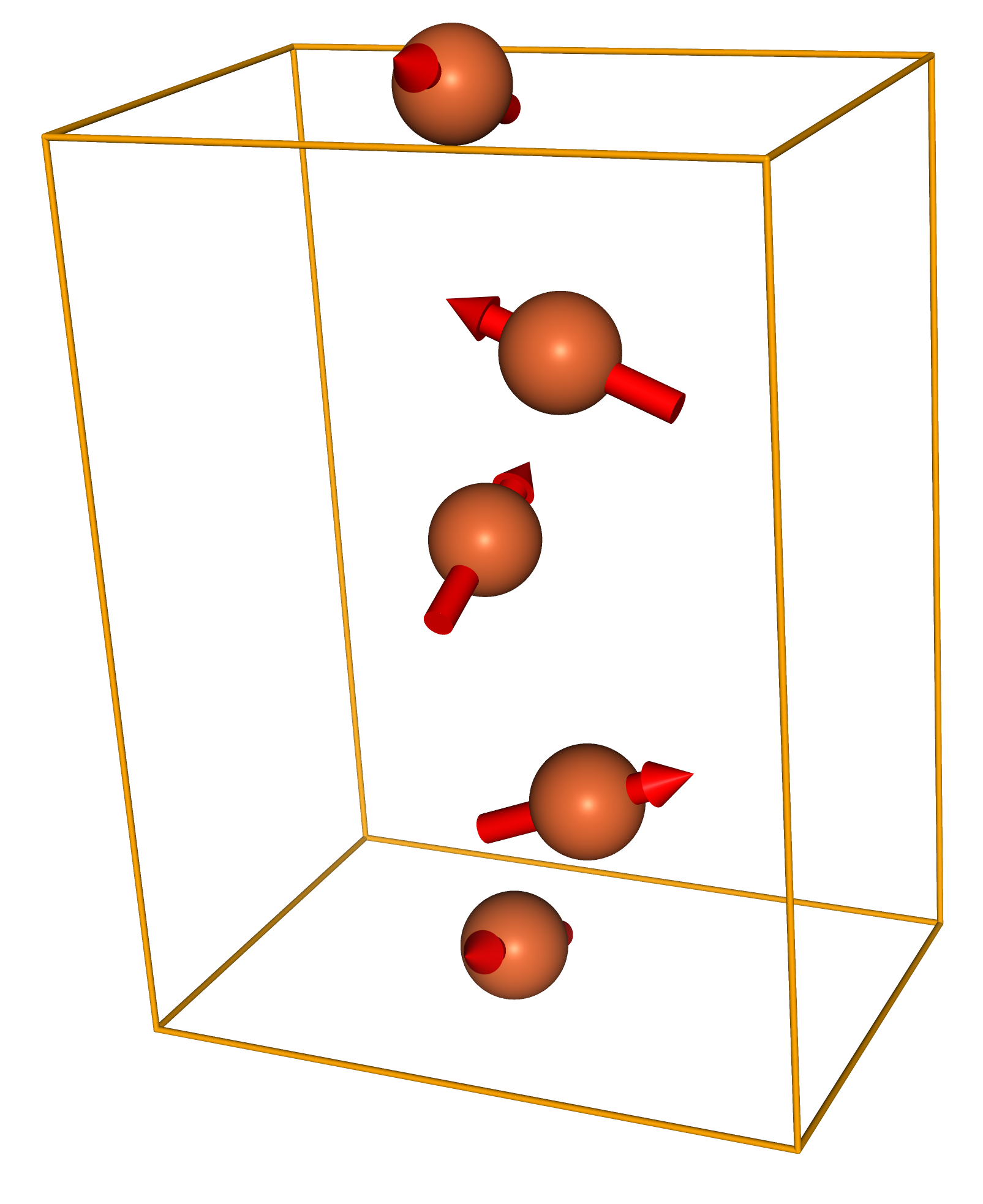}
  \caption{
    Spin arrangement example in Appendix~\ref{appx:spin-point-group}.
    Its spin point group is not closed under $\mathrm{O}(3) \times \mathrm{O}(3)$ due to the spin translation group.
  }
  \label{fig:counterexample_spin_point_group}
\end{figure}

In general, we cannot choose $\bm{W}_{\bm{R}}$ so that $\{ (\bm{R}, \bm{W}_{\bm{R}}) \}$ is a group under the multiplication in $\mathrm{O}(3) \times \mathrm{O}(3)$.
We show one of the counterexamples in which $\{ (\bm{R}, \bm{W}_{\bm{R}}) \}$ is not closed as a group, with a spin arrangement
\begin{align*}
    \bm{A}
        &= \begin{pmatrix}
            4 & 0 & 0 \\
            0 & 6 & 0 \\
            0 & 0 & 8 \\
        \end{pmatrix} \\
    \bm{X}
        &= \begin{pmatrix}
            0.4 & 0.6 & 0.4 & 0.4 \\
            0.4 & 0.6 & 0.4 & 0.4 \\
            0 & 0.25 & 0.5 & 0.75 \\
        \end{pmatrix} \\
    \bm{M}
        &= \begin{pmatrix}
            1 & 0 & -1 & 0 \\
            0 & 1 & 0 & -1 \\
            0.4 & 0.4 & 0.4 & 0.4 \\
        \end{pmatrix},
\end{align*}
illustrated in Fig.~\ref{fig:counterexample_spin_point_group}.
This spin arrangement is noncoplanar with $\mathcal{P}_{\mathrm{so}}(\mathcal{G}) = 1$.
The spin space group $\mathcal{G}$ is obtained as
\begin{align*}
    \mathcal{G}
        &= \left\{ g_{1}, g_{2}, g_{3}, g_{4}, g_{5}, g_{6}, g_{7}, g_{8} \right\} \left( \mathcal{T}(\mathcal{D}(\mathcal{G})) \times \mathcal{P}_{\mathrm{so}}(\mathcal{G}) \right) \\
    g_{1} &= \left( \left(1,            \bm{0}            \right)_{\bm{A}}, 1                 \right) \\
    g_{2} &= \left( \left(1,            \frac{1}{2}\bm{c} \right)_{\bm{A}}, 2_{z}             \right) \\
    g_{3} &= \left( \left(\overline{1}, \frac{3}{4}\bm{c} \right)_{\bm{A}}, m_{xy}            \right) \\
    g_{4} &= \left( \left(\overline{1}, \frac{1}{4}\bm{c} \right)_{\bm{A}}, m_{x\overline{y}} \right) \\
    g_{5} &= \left( \left(2,            \frac{3}{4}\bm{c} \right)_{\bm{A}}, 4^{-}_{z}         \right) \\
    g_{6} &= \left( \left(2,            \frac{1}{4}\bm{c} \right)_{\bm{A}}, 4^{+}_{z}         \right) \\
    g_{7} &= \left( \left(m,            \frac{1}{2}\bm{c} \right)_{\bm{A}}, m_{x}             \right) \\
    g_{8} &= \left( \left(m,            \bm{0}            \right)_{\bm{A}}, m_{y}             \right),
\end{align*}
where we denote a mirror operation along $xy$ axis as $m_{xy}$.
The spin point group of $\mathcal{G}$ is
\begin{align}
    \label{eq:spin-point-group-counterexample}
    \mathcal{U}(\mathcal{G})
        &= \left\{ (1, 1), (\overline{1}, m_{xy}), (2, 4^{-}_{z}), (m, m_{x}) \right\} \left( 1 \times \mathcal{B}(\mathcal{G}_{\mathrm{st}}(\mathcal{G})) \right) \\
    \mathcal{B}(\mathcal{G}_{\mathrm{st}}(\mathcal{G}))
        &= \left\{ 1, 2_{z} \right\} \nonumber
\end{align}
The coset representatives in Eq.~\eqref{eq:spin-point-group-counterexample} are not closed as a subgroup of $\mathrm{O}(3) \times \mathrm{O}(3)$ because $(2, 4^{-}_{z})^{-1} = (2, 4^{+}_{z})$ does not belong to the coset representatives.
In fact, we cannot choose coset representatives to form a subgroup of $\mathrm{O}(3) \times \mathrm{O}(3)$ in this example because $\mathcal{U}(\mathcal{G}) \cong \mathbb{Z}_{4}$, $\mathcal{B}(\mathcal{G}_{\mathrm{st}}(\mathcal{G})) \cong \mathbb{Z}_{2}$, and $\mathbb{Z}_{4}$ cannot be written as an internal semidirect product $\mathbb{Z}_{2} \rtimes \mathbb{Z}_{2}$.



\section{\label{appx:lattice}Generating transformation matrix from centering vectors}

To demonstrate how to generate a transformation matrix $\bm{V}$ in Sec.~\ref{sec:search:translation-subgroup}, we consider the following coplanar fcc structure with the conventional basis,
\begin{align*}
    \bm{A} &= \begin{pmatrix}
        a & 0 & 0 \\
        0 & a & 0 \\
        0 & 0 & a \\
    \end{pmatrix} \\
    \bm{X} &= \begin{pmatrix}
        0 & 0 & \frac{1}{2} & \frac{1}{2} \\
        0 & \frac{1}{2} & 0 & \frac{1}{2} \\
        0 & \frac{1}{2} & \frac{1}{2} & 0 \\
    \end{pmatrix} \\
    \bm{M} &= \begin{pmatrix}
        0 & 0 & m & m \\
        0 & 0 & 0 & 0 \\
        m & m & 0 & 0 \\
    \end{pmatrix}.
\end{align*}
One of the primitive basis vectors for the nonmagnetic crystal structure is
\begin{align*}
    \bm{A}_{\mathcal{S}}
        &= \bm{A} \bm{U}^{-1}
        = \frac{a}{2} \begin{pmatrix}
            0 & 1 & 1 \\
            1 & 0 & 1 \\
            1 & 1 & 0 \\
        \end{pmatrix} \\
    \bm{U}
        &= \begin{pmatrix}
            -1 & 1 & 1 \\
            1 & -1 & 1 \\
            1 & 1 & -1 \\
        \end{pmatrix}.
\end{align*}
There are four centering vectors in $\mathcal{T}_{\bm{A}_{\mathcal{S}}} / \mathcal{T}_{\bm{A}}$,
\begin{align*}
    \mathcal{T}_{\bm{A}_{\mathcal{S}}}
    &= g_{1} \mathcal{T}_{\bm{A}}
        \sqcup g_{2} \mathcal{T}_{\bm{A}}
        \sqcup g_{3} \mathcal{T}_{\bm{A}}
        \sqcup g_{4} \mathcal{T}_{\bm{A}} \\
    g_{1} &= \left( \bm{E}, (0, 0, 0)^{\top} \right)_{\bm{A}_{\mathcal{S}}} \\
    g_{2} &= \left( \bm{E}, (1, 0, 0)^{\top} \right)_{\bm{A}_{\mathcal{S}}} \\
    g_{3} &= \left( \bm{E}, (0, 1, 0)^{\top} \right)_{\bm{A}_{\mathcal{S}}} \\
    g_{4} &= \left( \bm{E}, (0, 0, 1)^{\top} \right)_{\bm{A}_{\mathcal{S}}}.
\end{align*}
The half of centering vectors form the translation subgroup of $\mathcal{D}(\mathcal{G})$,
\begin{align*}
    \mathcal{T}(\mathcal{D}(\mathcal{G}))
    = g_{1} \mathcal{T}_{\bm{A}} \sqcup g_{2} \mathcal{T}_{\bm{A}}.
\end{align*}
Because $\mathcal{T}(\mathcal{D}(\mathcal{G}))$ is spanned by column vectors of $\bm{U}$ and centering vectors of $g_{1}$ and $g_{2}$, we can rewrite $\mathcal{T}(\mathcal{D}(\mathcal{G}))$ as
\begin{align*}
    \mathcal{T}(\mathcal{D}(\mathcal{G}))
        &= \bm{A}_{\mathcal{S}} \tilde{\bm{U}} \mathbb{Z}^{5} \\
    \tilde{\bm{U}}
        &= \begin{pmatrix}
                -1 & 1 & 1 & 0 & 1 \\
                1 & -1 & 1 & 0 & 0 \\
                1 & 1 & -1 & 0 & 0 \\
            \end{pmatrix}.
\end{align*}

We need to find a transformation matrix $\bm{V} \in \mathbb{Z}^{3 \times 3}$ such that
\begin{align}
    \mathcal{T}(\mathcal{D}(\mathcal{G}))
        = \bm{A}_{\mathcal{S}} \bm{V} \mathbb{Z}^{3},
\end{align}
and it is known to be achieved by the Hermite normal form of $\tilde{\bm{U}}$ \cite{Cohen1993},
\begin{align*}
    \tilde{\bm{U}}
        = \begin{pmatrix}
            1 & 0 & 0 & 0 & 0 \\
            0 & 1 & 0 & 0 & 0 \\
            0 & 1 & 2 & 0 & 0 \\
        \end{pmatrix}
        \begin{pmatrix}
            -1 & 1  & 1  & 0 & 1 \\
            1  & -1 & 1  & 0 & 0 \\
            0  & 1  & -1 & 0 & 0 \\
            0  & 0  &  0 & 1 & 0 \\
            0  & 1  & 0  & 0 & 0 \\
        \end{pmatrix},
\end{align*}
where $\tilde{\bm{U}}$ is decomposed into a product of a lower triangular integer matrix and a unimodular matrix.
Because the above $5 \times 5$ integer matrix is unimodular, we obtain
\begin{align*}
    \mathcal{T}(\mathcal{D}(\mathcal{G}))
        &= \bm{A}_{\mathcal{S}}
            \begin{pmatrix}
                1 & 0 & 0 & 0 & 0 \\
                0 & 1 & 0 & 0 & 0 \\
                0 & 1 & 2 & 0 & 0 \\
            \end{pmatrix} \mathbb{Z}^{5} \\
        &= \bm{A}_{\mathcal{S}}
            \begin{pmatrix}
                1 & 0 & 0 \\
                0 & 1 & 0 \\
                0 & 1 & 2 \\
            \end{pmatrix}
            \mathbb{Z}^{3}.
\end{align*}
Consequently, we generate the integer matrix
\begin{align*}
    \bm{V}
        = \begin{pmatrix}
            1 & 0 & 0 \\
            0 & 1 & 0 \\
            0 & 1 & 2
        \end{pmatrix}.
\end{align*}





\ack{}


\referencelist[references] 


\begin{table}
    \caption{Notation and terminology in this paper.}
    \label{tab:terminology}
    \small
    \begin{tabularx}{\textwidth}{ll}
        Symbol & Meaning \\ \hline
        $A \sqcup B$ & Disjoint union of set $A$ and $B$ with $A \cap B = \varnothing$ \\
        $\mathrm{E}(3)$ & Three-dimensional Euclidean group \\
        $\mathrm{O}(3)$ & Three-dimensional orthogonal group \\
        $\bm{E}$ & Identity matrix \\
        $\mathfrak{S}_{N}$ & Symmetric group of degree $N$ \\
        $\norm{\bm{y}}_{2} = \sqrt{ \bm{y}^{\top}\bm{y} }$ & $l^{2}$ norm of vector $\bm{y}$ \\
        $\norm{\bm{B}}_{F} = \sqrt{\Tr \bm{B}^{\top}\bm{B}} $ & Frobenius norm of matrix $\bm{B}$ \\
        $1$ & Trivial group \\
        \hline
        $\bm{r}$ & Position in Cartesian coordinates \\
        $\bm{m}$ & Magnetic moments \\
        Spin symmetry operation $(g, \bm{W})$ & Pair of spatial operation $g$ and spin rotation $\bm{W}$ \\
        $( \overline{\bm{R}}, \overline{\bm{v}} )_{\bm{A}} = (\bm{A} \overline{\bm{R}} \bm{A}^{-1}, \bm{A} \overline{\bm{v}})$ & Spatial operation with basis vectors $\bm{A}$ \\
        \hline
        $\bm{A} = (\bm{a}_{1}, \bm{a}_{2}, \bm{a}_{3})$ & Basis vectors \\
        $\bm{X} = (\bm{x}_{1}, \cdots, \bm{x}_{N})$ & Array of point coordinates \\
        $\bm{T} = (t_{1}, \cdots, t_{N})$ & Array of atomic types \\
        $\bm{M} = (\bm{m}_{1}, \cdots, \bm{m}_{N})$ & Array of magnetic moments \\
        Spin arrangement $(\bm{A}, \bm{X}, \bm{T}, \bm{M})$ & Pair of crystal structure and magnetic moments \\
        $(\bm{A}, \bm{X}, \bm{T})$ & Crystal structure ignoring magnetic moments \\
        $\epsilon_{\mathrm{mag}}$ & Absolute tolerance to compare magnetic moments \\
        \hline
        $\mathcal{S}, \mathcal{R}$ & Space group\\
        Translation subgroup $\mathcal{T}(\mathcal{R})$ & Translation parts of $\mathcal{R}$ with identity rotations \\
        Point group $\mathcal{P}(\mathcal{R})$ & Group obtained from the rotation parts of $\mathcal{R} / \mathcal{T}(\mathcal{R})$ \\
        $\mathcal{T}_{\bm{A}}$ & Translation group spanned by $\bm{A}$ \\
        $\sigma_{g}$ & Permutation of sites induced by $g$ \\
        $\bm{A}_{\mathcal{S}} = \bm{A} \bm{U}^{-1}$ & Primitive basis vectors of $\mathcal{T}(\mathcal{S})$ \\
        \hline
        Spin-only group $\mathcal{P}_{\mathrm{so}} = \mathcal{P}_{\mathrm{so}}(\mathcal{G})$ & Spin-rotation parts with identity spatial operations \\
        $\sigma_{i}, \hat{\bm{n}_{i}}$ & Eigenvalue and eigenvector of $\bm{M}\bm{M}^{\top}$ \\
        $\hat{\bm{n}}_{\parallel}$ & Parallel direction for collinear spin arrangement \\
        $\hat{\bm{n}_{\perp}}$ & Perpendicular direction for coplanar spin arrangement \\
        \hline
        $\bm{A}_{\mathcal{D}} = \bm{A}_{\mathcal{S}} \bm{V}$ & Basis vectors of $\mathcal{T}(\mathcal{D}(\mathcal{G}))$ \\
        \hline
        Spin translation group $\mathcal{G}_{\mathrm{st}} = \mathcal{G}_{\mathrm{st}}(\mathcal{G})$ & Subgroup of $\mathcal{G}$ with identity rotations \\
        $\bm{M}_{g} = \left( \bm{m}_{\sigma_{g}(1)} \cdots \bm{m}_{\sigma_{g}(N)} \right)$ & Array of magnetic moments permuted by $\sigma_{g}$ \\
        \hline
        Spin space group $\mathcal{G}$ & See Sec.~\ref{sec:gs:spin-space-group} \\
        Family space group $\mathcal{F}(\mathcal{G})$ & Space group composed of spatial parts of $\mathcal{G}$ \\
        Maximal space subgroup $\mathcal{D} = \mathcal{D}(\mathcal{G})$ & Subgroup of $\mathcal{G}$ with identity spin-rotation parts \\
        Spin-space-group type & See Sec.~\ref{sec:gs:spin-space-group} \\
        $(\bm{P}, \bm{p})$ & Transformation on spatial coordinates \\
        $\bm{Q}$ & Transformation matrix on spin coordinates \\
        \hline
        Family spin point group $\mathcal{B}(\mathcal{G})$ & Spin-rotation parts of $\mathcal{G}$ \\
        Spin point group $\mathcal{U}(\mathcal{G})$ & Pairs of spatial and spin rotations of $\mathcal{G}$ \\
        \hline
        $\mathcal{M}$ & Magnetic space group \\
    \end{tabularx}
\end{table}

\begin{table}
  \caption{
    Classification of spin-only groups up to transformations.
    The spin-only groups are represented in Hellmann--Mauguin symbols \cite{hahn2016point}.
  }
  \label{tab:spin_only_group}
  \begin{tabularx}{\textwidth}{sss}
    Spin arrangement & Spin-only group & Eigenvalues of $\bm{M}\bm{M}^{\top}$ \\
    \hline
    Nonmagnetic & $\infty \infty m \cong \mathrm{O}(3)$                  & $\sigma_{1} = \sigma_{2} = \sigma_{3} = 0$ \\
    Collinear   & $\infty m \cong \mathrm{SO}(2) \rtimes \mathbb{Z}_{2}$ & $\sigma_{1} > \sigma_{2} = \sigma_{3} = 0$ \\
    Coplanar    & $m$                                                    & $\sigma_{1} \geq \sigma_{2} > \sigma_{3} = 0$ \\
    Noncoplanar & $1$                                                    & $\sigma_{1} \geq \sigma_{2}  \geq \sigma_{3} > 0$ \\
  \end{tabularx}
\end{table}

@book{Opechowski1986,
  author        = {W. Opechowski},
  title         = {Crystallographic and Metacrystallographic Groups},
  publisher     = {North-Holland},
  year          = {1986}
}

@article{LITVIN1974538,
  title         = {Spin groups},
  journal       = {Physica},
  volume        = {76},
  number        = {3},
  pages         = {538--554},
  year          = {1974},
  issn          = {0031-8914},
  doi           = {https://doi.org/10.1016/0031-8914(74)90157-8},
  author        = {D.B. Litvin and W. Opechowski}
}

@article{doi:10.1063/1.1708514,
  author        = {Brinkman,W.  and Elliott,R. J.},
  title         = {Space Group Theory for Spin Waves},
  journal       = {J. Appl. Phys.},
  volume        = {37},
  number        = {3},
  pages         = {1457--1459},
  year          = {1966},
  doi           = {10.1063/1.1708514}
}

@article{doi:10.1098/rspa.1966.0211,
  author        = {Brinkman, W. F.  and Elliott, Roger James  and Peierls, Rudolf Ernst},
  title         = {Theory of spin-space groups},
  journal       = {Proc. Math. Phys. Eng. Sci.},
  volume        = {294},
  number        = {1438},
  pages         = {343--358},
  year          = {1966},
  doi           = {10.1098/rspa.1966.0211}
}

@article{PhysRevX.12.021016,
  title         = {Spin-Group Symmetry in Magnetic Materials with Negligible Spin-Orbit Coupling},
  author        = {Liu, Pengfei and Li, Jiayu and Han, Jingzhi and Wan, Xiangang and Liu, Qihang},
  journal       = {Phys. Rev. X},
  volume        = {12},
  issue         = {2},
  pages         = {021016},
  numpages      = {19},
  year          = {2022},
  month         = {Apr},
  publisher     = {American Physical Society},
  doi           = {10.1103/PhysRevX.12.021016}
}

@article{Litvin:a09793,
  author        = "Litvin, D. B.",
  title         = "{Spin translation groups and neutron diffraction analysis}",
  journal       = "Acta Cryst. A",
  year          = "1973",
  volume        = "29",
  number        = "6",
  pages         = "651--660",
  month         = "Nov",
  doi           = {10.1107/S0567739473001658}
}

@article{Litvin:a14103,
  author        = "Litvin, D. B.",
  title         = "{Spin point groups}",
  journal       = "Acta Cryst. A",
  year          = "1977",
  volume        = "33",
  number        = "2",
  pages         = "279--287",
  month         = "Mar",
  doi           = {10.1107/S0567739477000709}
}

@book{10.1093/acprof:oso/9780198510581.001.0001,
  author        = {Gower, John C and Dijksterhuis, Garmt B},
  title         = "Procrustes Problems",
  publisher     = {Oxford University Press},
  year          = {2004},
  month         = {01},
  isbn          = {9780198510581},
  doi           = {10.1093/acprof:oso/9780198510581.001.0001}
}

@article{PhysRevB.106.144433,
  title         = {Magnetic interactions in AB-stacked kagome lattices: Magnetic structure, symmetry, and duality},
  author        = {Zelenskiy, A. and Monchesky, T. L. and Plumer, M. L. and Southern, B. W.},
  journal       = {Phys. Rev. B},
  volume        = {106},
  issue         = {14},
  pages         = {144433},
  numpages      = {22},
  year          = {2022},
  month         = {Oct},
  publisher     = {American Physical Society},
  doi           = {10.1103/PhysRevB.106.144433}
}

@article{PhysRev.122.1402,
  title         = {Magnetic Structure of Chromium Selenide},
  author        = {Corliss, L. M. and Elliott, N. and Hastings, J. M. and Sass, R. L.},
  journal       = {Phys. Rev.},
  volume        = {122},
  issue         = {5},
  pages         = {1402--1406},
  numpages      = {0},
  year          = {1961},
  month         = {Jun},
  publisher     = {American Physical Society},
  doi           = {10.1103/PhysRev.122.1402}
}

@article{Gallego:ks5532,
  author        = "Gallego, Samuel V. and Perez-Mato, J. Manuel and Elcoro, Luis and Tasci, Emre S. and Hanson, Robert M. and Momma, Koichi and Aroyo, Mois I. and Madariaga, Gotzon",
  title         = "{{\it MAGNDATA}: towards a database of magnetic structures. I.The commensurate case}",
  journal       = "J. Appl. Cryst.",
  year          = "2016",
  volume        = "49",
  number        = "5",
  pages         = "1750--1776",
  month         = "Oct",
  doi           = {10.1107/S1600576716012863}
}

@book{litvin2014magnetic,
  title         = {Magnetic group tables},
  author        = {Litvin, Daniel B},
  publisher     = {IUCr},
  note          = {\url{https://www.iucr.org/publ/978-0-9553602-2-0}},
  year          = {2014}
}

@misc{spglibv1,
  author        = {Atsushi Togo and Isao Tanaka},
  title         = {$\texttt{Spglib}$: a software library for crystal symmetry search},
  eprint        = {arXiv:1808.01590},
  howpublished  = {\url{https://github.com/spglib/spglib}},
  year          = {2018}
}

@article{spglibv2,
  author        = {Kohei Shinohara and Atsushi Togo and Isao Tanaka},
  title         = {Algorithms for magnetic symmetry operation search and identification of magnetic space group from magnetic crystal structure},
  journal       = {Acta Cryst. A},
  eprint        = {arXiv:2211.15008},
  howpublished  = {\url{https://arxiv.org/abs/2211.15008}},
  year          = {to be published 2023},
}

@book{Cohen1993,
  doi           = {10.1007/978-3-662-02945-9},
  year          = {1993},
  publisher     = {Springer Berlin Heidelberg},
  author        = {Henri Cohen},
  title         = {A Course in Computational Algebraic Number Theory}
}

@incollection{nebe20061,
  title         = {The mathematical background of the subgroup tables},
  author        = {Nebe, Gabriele},
  chapter       = {1.4},
  pages         = {27--40},
  booktitle     = {International Tables for Crystallography},
  editor        = {H. Wondratschek and U. M\"{u}ller},
  edition       = {2},
  volume        = {A1},
  year          = {2011},
  publisher     = {Springer}
}

@article{Stokes:vk5013,
  author        = "Stokes, Harold T. and Campbell, Branton J.",
  title         = "{A general algorithm for generating isotropy subgroups in superspace}",
  journal       = "Acta Cryst. A",
  year          = "2017",
  volume        = "73",
  number        = "1",
  pages         = "4--13",
  month         = "Jan",
  doi           = {10.1107/S2053273316017629}
}

@article{PhysRevLett.119.187204,
  title         = {Spin-Polarized Current in Noncollinear Antiferromagnets},
  author        = {\ifmmode \check{Z}\else \v{Z}\fi{}elezn\'y, Jakub and Zhang, Yang and Felser, Claudia and Yan, Binghai},
  journal       = {Phys. Rev. Lett.},
  volume        = {119},
  issue         = {18},
  pages         = {187204},
  numpages      = {7},
  year          = {2017},
  month         = {Nov},
  publisher     = {American Physical Society},
  doi           = {10.1103/PhysRevLett.119.187204}
}

@article{Zhang2018,
  doi           = {10.1088/1367-2630/aad1eb},
  year          = {2018},
  month         = jul,
  publisher     = {{IOP} Publishing},
  volume        = {20},
  number        = {7},
  pages         = {073028},
  author        = {Yang Zhang and Jakub {\v{Z}}elezn{\'{y}} and Yan Sun and Jeroen van den Brink and Binghai Yan},
  title         = {Spin Hall effect emerging from a noncollinear magnetic lattice without spin{\textendash}orbit coupling},
  journal       = {New J. Phys.}
}

@article{PhysRevX.12.040501,
  title         = {Emerging Research Landscape of Altermagnetism},
  author        = {\ifmmode \check{S}\else \v{S}\fi{}mejkal, Libor and Sinova, Jairo and Jungwirth, Tomas},
  journal       = {Phys. Rev. X},
  volume        = {12},
  issue         = {4},
  pages         = {040501},
  numpages      = {27},
  year          = {2022},
  month         = {Dec},
  publisher     = {American Physical Society},
  doi           = {10.1103/PhysRevX.12.040501}
}

@article{PhysRevB.105.064430,
  title         = {Spin-space groups and magnon band topology},
  author        = {Corticelli, A. and Moessner, R. and McClarty, P. A.},
  journal       = {Phys. Rev. B},
  volume        = {105},
  issue         = {6},
  pages         = {064430},
  numpages      = {28},
  year          = {2022},
  month         = {Feb},
  publisher     = {American Physical Society},
  doi           = {10.1103/PhysRevB.105.064430}
}

@article{PhysRevX.12.031042,
  title         = {Beyond Conventional Ferromagnetism and Antiferromagnetism: A Phase with Nonrelativistic Spin and Crystal Rotation Symmetry},
  author        = {\ifmmode \check{S}\else \v{S}\fi{}mejkal, Libor and Sinova, Jairo and Jungwirth, Tomas},
  journal       = {Phys. Rev. X},
  volume        = {12},
  issue         = {3},
  pages         = {031042},
  numpages      = {16},
  year          = {2022},
  month         = {Sep},
  publisher     = {American Physical Society},
  doi           = {10.1103/PhysRevX.12.031042}
}

@misc{spinspg-application,
  author        = {Watanabe, Hikaru and Shinohara, Kohei and Nomoto, Takuya and Togo, Atsushi and Arita Ryotaro},
  title         = {Spin-group symmetry analysis of emergent electromagnetic properties},
  eprint        = {},
  howpublished  = {submitted},
  year          = {2023}
}

@incollection{hahn2016point,
  title         = {Point groups and crystal classes},
  author        = {Hahn, Th and Klapper, H and M{\"u}ller, U and Aroyo, MI},
  chapter       = {3.2},
  pages         = {720--776},
  booktitle     = {International Tables for Crystallography},
  editor        = {M. I. Aroyo},
  edition       = {6},
  volume        = {A},
  year          = {2016},
  publisher     = {Wiley}
}

@article{PhysRevB.77.224115,
  title = {Algorithm for generating derivative structures},
  author = {Hart, Gus L. W. and Forcade, Rodney W.},
  journal = {Phys. Rev. B},
  volume = {77},
  issue = {22},
  pages = {224115},
  numpages = {12},
  year = {2008},
  month = {Jun},
  publisher = {American Physical Society},
  doi = {10.1103/PhysRevB.77.224115},
  url = {https://link.aps.org/doi/10.1103/PhysRevB.77.224115}
}

@misc{2307.10364,
  Author = {Zhenyu Xiao and Jianzhou Zhao and Yanqi Li and Ryuichi Shindou and Zhi-Da Song},
  Title = {Spin Space Groups: Full Classification and Applications},
  Year = {2023},
  Eprint = {arXiv:2307.10364},
  howpublished={\url{https://arxiv.org/abs/2307.10364}}
}

@misc{2307.10369,
  Author = {Jun Ren and Xiaobing Chen and Yanzhou Zhu and Yutong Yu and Ao Zhang and Jiayu Li and Caiheng Li and Qihang Liu},
  Title = {Enumeration and representation of spin space groups},
  Year = {2023},
  Eprint = {arXiv:2307.10369},
  howpublished={\url{https://arxiv.org/abs/2307.10369}}
}

@misc{2307.10371,
  Author = {Yi Jiang and Ziyin Song and Tiannian Zhu and Zhong Fang and Hongming Weng and Zheng-Xin Liu and Jian Yang and Chen Fang},
  Title = {Enumeration of spin-space groups: Towards a complete description of symmetries of magnetic orders},
  Year = {2023},
  Eprint = {arXiv:2307.10371},
  howpublished={\url{https://arxiv.org/abs/2307.10371}}
}

@misc{2105.12738,
  Author = {Jian Yang and Zheng-Xin Liu and Chen Fang},
  Title = {Symmetry invariants in magnetically ordered systems having weak spin-orbit coupling},
  Year = {2021},
  Eprint = {arXiv:2105.12738},
  howpublished={\url{https://arxiv.org/abs/2105.12738}}
}
\end{document}